\documentclass[12pt]{article}
\usepackage{graphicx,xcolor,booktabs}
\usepackage[colorlinks,linkcolor=blue,citecolor=blue,urlcolor=blue,bookmarks,bookmarksnumbered]{hyperref}
\usepackage{amsmath,amssymb,amsfonts,mathrsfs}
\usepackage[paper=letterpaper, margin=1.0in, top=1.5in, bottom=0.33in]{geometry}
\usepackage{cite}
\RequirePackage{tikz}
 \usetikzlibrary{arrows}
 
\usepackage{enumitem}

\newcommand{\be}{\begin{equation}}
\newcommand{\ee}{\end{equation}}
\newcommand{\bea}{\begin{eqnarray}}
\newcommand{\eea}{\end{eqnarray}}
\newcommand{\ba}{\begin{array}}
\newcommand{\ea}{\end{array}}
\newcommand{\ben}{\begin{enumerate}}
\newcommand{\een}{\end{enumerate}}
\newcommand{\bi}{\begin{itemize}}
\newcommand{\ei}{\end{itemize}}
\newcommand{\bc}{\begin{center}}
\newcommand{\ec}{\end{center}}
\newcommand{\bfig}{\begin{figure}}
\newcommand{\efig}{\end{figure}}
\newcommand{\bq}{\begin{quotation}}
\newcommand{\eq}{\end{quotation}}
\newcommand{\bt}{\begin{table}}
\newcommand{\et}{\end{table}}
\newcommand{\btab}{\begin{tabular}}
\newcommand{\etab}{\end{tabular}}
\newcommand{\bs}{\begin{slide}}
\newcommand{\es}{\end{slide}}

\newcommand{\pa}{\partial}

\newcommand{\X}{\mathbb{X}}
\newcommand{\cX}{\mathcal{X}}

\newcommand{\bra}{\langle}
\newcommand{\ket}{\rangle}

\newcommand{\vev}[1]{\langle #1 \rangle}

\newdimen\lft\lft=0pt

\newcommand{\beq}{\begin{eqnarray}}
\newcommand{\eeq}{\end{eqnarray}}
\newcommand{\beqn}{\begin{eqnarray}}
\newcommand{\eeqn}{\end{eqnarray}}

\def\pa{\partial}
\newcommand{\rd}{\mathrm{d}}
\newcommand{\rD}{\mathrm{D}}
\newcommand{\ls}{\ell_s}

\let\SS=\S % LaTeX's Section symbol
\def\S{\mathbb{S}}

\def\bra{\langle}
\def\ket{\rangle}

\def\lL{l_\Lambda}

\let\ba=\overline

\let\d=\delta

\def\define{\buildrel{\rm def}\over=}

\let\j=\psi

\let\L=\Lambda

\def\Lcc{\Lambda_{\rm cc}}

\let\p=\pi
\def\Seff{S_{\rm eff}}
\let\t=\tau

\def\vev#1{\left\langle#1\right\rangle}

\def\RR{\relax\leavevmode
       \ifmmode\mathchoice
       {\hbox{\sf R\kern-.4em R}}
       {\hbox{\sf R\kern-.4em R}}
       {\lower.9pt\hbox{\scriptsize\sf R\kern-.36em R}}
       {\lower1.2pt\hbox{\tiny\sf R\kern-.36em R}}
       \else{\sf R\kern-.4em R}\fi}

\def\resetby#1#2{\@addtoreset{#2}{#1}}
\def\seceq{\@addtoreset{equation}{section}
              \def\theequation{\thesection.\arabic{equation}}}

\def\Label#1{\label{#1}%
                \smash{\hbox to0pt{\raise1ex\hbox{\tiny[#1]}\hss}}}
\def\noLabels{\let\Label=\label}

\DeclareMathOperator{\Tr}{\textrm{Tr}}

\let\sss=\scriptscriptstyle

\def\GeV{\text{GeV}}
\def\MeV{\text{MeV}}
\def\keV{\text{keV}}
\def\eV{\text{eV}}
\def\xX{\xi_{\mathcal{X}}}

\def\tx{{\mathord{\tilde x}}}

\def\2#1{{\color{blue}#1}}

 \allowdisplaybreaks
 \numberwithin{equation}{section}
 
 \parskip\medskipamount
\begin{document}
\renewcommand{\thefootnote}{\fnsymbol{footnote}}
\pagestyle{empty}

\begin{center}

\vskip 10mm
\begin{center}\Large \bf
    Quantum Gravity as Gravitized Quantum Theory%
    \footnote{Prepared for the Proceedings of the 3rd Conference on Nonlinearity (September, 2023), held at the Serbian Academy of Nonlinear Sciences, Belgrade, Serbia}
\end{center}
\vskip 10mm

\centerline{{\bf
Tristan H\"{u}bsch${}^{1}$\footnote{\tt thubsch@howard.edu}
and
Djordje Minic${}^{2}$\footnote{\tt dminic@vt.edu}
}}

{\small\it
${}^1$Department of Physics and Astronomy, Howard University, Washington, D.C.  20059, U.S.A. \\
${}^2$Department  of Physics, Virginia Tech, Blacksburg, VA 24061, U.S.A. \\
${}$ \\
}

\end{center}

\vskip 5mm
{\em Dedicated to the memory of Joy Rosenthal (1966-2023)}
\vskip 15mm

\begin{abstract}
Starting from a new understanding of the vacuum energy problem based on the combination of the phase space regularization and the holographic bound, we argue that quantum gravity should be understood as {\em gravitized quantum theory,} that is, quantum theory wherein the geometry and topology of the state-space if fully dynamical, in analogy with the dynamical nature of spacetime in Einstein's general relativity. Apart from the vacuum energy problem viewed as a quantum gravity problem, we discuss the ``smoking gun'' experiments involving higher order quantum interference, as well as experimental probes of the statistics of spacetime quanta. Finally, we address the conundrum of the intricately patterned spectrum of masses of elementary particles as well as their mixing angles, as another telltale problem of quantum gravity viewed as gravitized quantum theory.
\end{abstract}
\vfill

%%%%%%%%%%%%%%%%%%%%%%%%%%%%%%%%%%%%%%%%%%%%%%%%
%: Table of Contents
\begingroup
\baselineskip=10pt %\smallskipamount
\setcounter{tocdepth}{3}
  \begin{center}
    \begin{minipage}{140mm}
      \tableofcontents
    \end{minipage}
  \end{center}
\endgroup

\setcounter{footnote}{0}
\renewcommand{\thefootnote}{\arabic{footnote}}
\vfill
\clearpage
\pagestyle{plain}

%%%%%%%%%%%%%%%%%%%%%%%%%%%%%%%%%%%%%%%%%%%%%%%%
%: The Paper
\setcounter{page}{1}
\section{Introduction: Why Quantum Gravity?}
\label{s:IRS}
A final, comprehensive, cohesive and consistent unification of quantum physics and gravi\-ta\-tion (relativistic as well non-relativistic) --- dubbed quantum gravity (QG) --- has been a coveted goal for the better part of the past century. Yet, even the very concept of {\em what,} exactly, QG ought to be (in fact, even whether any such a thing can exist) continues to be hotly debated (to say the least\cite{Oriti:2009App, deBoer:2022zka}, and references therein), especially in the absence of any guiding and illuminating sharp empirical facts.

Indicative of its profoundly vexing difficulty, over two dozen theoretical approaches to QG (see\cite{Oriti:2009App, deBoer:2022zka} and references therein) have been developed by now:\footnote{Even in a review such as this, there can be no hope for any semblance of doing justice in reviewing these approaches and important questions, so we list the most prominent among them to indicate the breadth of the field, and to help the interested reader with starting their own searching efforts (see also, for example, the ``International Society for Quantum Gravity'' YouTube channel,
\url{https://www.youtube.com/@isqg423}, %\url{https://www.youtube.com/@isqg423/videos},
and the lecture
\url{https://www.youtube.com/watch?v=7T6pinf7fLg}\,).}
{\sl the effective field theory (EFT) approach, asymptotic safety, supergravity, string theory, holography (AdS/CFT and generalizations: 
de~Sitter (dS), celestial, corner symmetry), Euclidean quantum gravity, 
topological (and categorical) quantum field theory, canonical quantum gravity, loop quantum gravity (Hamiltonian/spin networks and covariant/spin foams), causal sets, quantum cosmology, group field theory, emergent gravity in condensed matter, Regge calculus, (causal) dynamical triangulation, non-commutative geometry, twistor theory, Horava-Lifshitz gravity, analog gravity models, modified/massive gravity, gauge theories of gravity, non-local theories of gravity,
shape dynamics, entropic gravity, quantum gravity phenomenology, quantum graphity, gravitized quantum theory, etc.}
There are also many key questions to be answered by quantum gravity, which may be grouped as follows.
{\sl\bfseries Questions about spacetime:}
{\sl resolution of singularities (black hole and cosmological), quantum black holes (including black hole (BH) entropy, information puzzle), 
astrophysics of the (resolved) BH singularity, cosmology of the (resolved) initial singularity, fine tuning of the initial state, early universe cosmology,
relevance of quantum gravity for structure formation,
quantum structure of spacetime and its phenomenology, the vacuum energy problem, etc.}
{\sl\bfseries Questions about matter:}
{\sl the observed hierarchy of scales,
origin of the Standard Models of particle physics and cosmology, masses and couplings of fundamental particles, dark matter and dark energy, baryogenesis, etc.}
{\sl\bfseries Conceptual questions:}
{\sl emergence of quantum theory, quantum/classical transition, quantum measurement problem, topology change,
problem of time, closed timelike curves,
emergence of spacetime,
origin of inertia, etc.} A tall order indeed!

Herein, we consider this difficult problem rather literally {\em from the ground up\/}: starting with the long-standing and oft-ridiculed as an impossible problem in effective field theory (EFT), the vacuum energy, i.e., the cosmological constant, $\Lcc$, which has been measured 
\cite{SupernovaSearchTeam:1998fmf, SupernovaCosmologyProject:1998vns},
and which could be considered as one established empirical fact of quantum gravity. In particular, we present a new computation of the vacuum energy fundamentally based on the short-distance/long-distance interplay between the phase space geometry and the holographic bound. 
Motivated by this new calculation of the vacuum energy that matches the observed value (interpreted as the cosmological constant in Einstein's gravitational equations, and in principle the {\em first measured quantum gravity observable\/}), we argue that the essential ingredient of quantum gravity is a dynamical quantum phase space, and thus a dynamical form of the Born rule used to define general quantum probabilities and observables of quantum gravity. This identifies {\em gravitization of quantum theory\/}\footnote{By ``gravitizing'' we mean rendering the geometry and even topology of the quantum theory {\em state-space} fully dynamical, akin to how spacetime becomes dynamical in Einstein's theory of gravity.} as the key feature by which the approach presented in this review differs from traditional approaches to quantum gravity. We present a general discussion of quantum spacetime rewriting of quantum theory and a dynamical quantum spacetime formulation found in a non-commutative and phase-space-like formulation of string theory, the metastring\cite{Freidel:2015pka}. The gravitization of quantum theory sheds light on the origin of quantum theory and quantum field theory and so gravitized quantum theory can be understood as a {\em metaquantum} theory. Also gravitized quantum theory in turn implies very specific experimental probes of higher order quantum interference processes which are impossible in local quantum field theory. Furthermore, the statistics of the ``atomic'' constituents of spacetime is illuminated in this context, and contrasted to the familiar spin-statistics relation established for the matter degrees of freedom. The statistics of spacetime quanta can be also empirically probed using gravitational interferometry.

The calculation of the vacuum energy can be extended to compute the masses (gravitational charges) of fundamental particles in the matter sector and find 
a cascade of seesaw-determined mass-scales, which befit the Standard Model (SM) extraordinarily well. This intricate pattern of SM fermion masses\cite{Weinberg:2020zba} and the vacuum energy problem\cite{Weinberg:1988cp} are the two most vexing issues in fundamental physics --- and so also of quantum gravity. 
Following\cite{Freidel:2022ryr, Berglund:2022qsb} and sharpening the recent analysis\cite{Berglund:2023gur}, we provide a conceptually more cohesive and coherent framework and show that its {\em unifying,} ultraviolet (UV)/infrared (IR) mixing solution to these two problems of fundamental physics also addresses the gauge hierarchy (Higgs mass) problem
as well as the problem of mixing both in the quark and neutrino sectors\cite{Berglund:2023gur, Minic:2023oty}.

In what follows we emphasize the quantum gravitational nature of these problems, represented by the explicit appearance of widely separated short-distance and long-distance scales. Thus quantum gravity is not a (purely) Planck scale phenomenon, as it finds its physical manifestations at
all scales via the spectrum of elementary particles.
Furthermore, low energy phenomena of higher order quantum interference as well as the experimental probes of the statistics of spacetime quanta, are indicative of both the infrared and ultraviolet nature of quantum gravity phenomenology.
From our viewpoint, one of the most pressing issues in the research in quantum gravity (where theory reigns supreme) is the development of various experimental probes, at various scales, of such quantum gravity phenomenology.

\begin{figure}[h!tb]
$$
\begin{tikzpicture}[xscale=.95,every node/.style={inner sep=0,outer sep=2pt}]
    \path[use as bounding box](0,0)--(16,-5.75);
    \node(CH) at ( 0, 0) {$M_{CH}$};
     \path(.1,-.4)node{\footnotesize$10^{-34}\,\eV$};
    \node(Pl) at (16, 0) {$M_P$};
     \path(16.35,-.4)node{\footnotesize$10^{28}\,\eV$};
    \node(ML) at (16*1/2,-1) {$M_\L$};
     \path(16*1/2-1.3,-1.15)node{\footnotesize$10^{-3}\,\eV$};
    \node(SM) at (16*13/14,-1.3) {$M_{SM}$};
    \path(16*13/14+.5,-1.7)node
    {\footnotesize$4.5{\times}10^{23}\,\eV$};
    \node(MH) at (16*3/4,-2) {$\2{M_H^\star}$};
    \node(MZ) at (16*2/3,-2) {$M_{BZ}$};
    \node(mt) at (16*3/4,-4) {$\{\2{m^\star_{t,b,\tau}}\}$};
    \node(mc) at (16*17/24,-5) {$\{m_{c,s,\mu}\}$};
    \node(mu) at (16*5/8,-6) {$\{m_{u,d,e}\}$};
    \node(m3) at (16*4/7,-4) {$m_{\nu_3}$};
    \node(m2) at (16*15/28,-5) {$m_{\nu_2}$};
    \node(m1) at (16*13/28,-6) {$m_{\nu_1}$};
    \draw[thick, ->](CH)--(ML); \draw[thick, ->](Pl)--(ML);
    \draw[thick, ->](ML)--(MH); \draw[thick, ->](Pl)--(MH);
    \draw[thick, ->](ML)--(MZ); \draw[thick, ->](Pl)--(MZ);
    \draw[thick, ->](CH)--(SM); \draw[thick, ->](Pl)--(SM);
    \draw[->](MH)--(mt);
    \draw[->](MZ)--(mc); \draw[->](mt)--(mc);
    \draw[->](MZ)--(mu); \draw[->](mc)--(mu);
    \draw[->](MH)--(m3); \draw[->](SM)--(m3);
    \draw[->](ML)--(m2); \draw[->](m3)--(m2);
    \draw[->](ML)--(m1); \draw[->](m2)--(m1);
    \color{purple}
    \node(Vcb) at (16*1/48+14,-6) {$V_{cb}$};
     \path(16*1/48+14.95,-6)node{\footnotesize($0.050$)};
    \node(Vtd) at (16*3/48+14,-5) {$V_{td}$};
     \path(16*3/48+14.95,-5)node{\footnotesize($0.011$)};
    \node(Vub) at (16*4/48+14,-4) {$V_{ub}$};
     \path(16*4/48+14.4,-3.6)node{\footnotesize($0.002$)};
    \node(Um3) at (16*1/56+2,-6) {$U_{\mu3}$};
     \path(16*1/56+1.1,-6)node{\footnotesize($0.50$)};
    \node(Ut1) at (16*3/56+2,-5) {$U_{\tau1}$};
     \path(16*3/56+1.1,-5)node{\footnotesize($0.13$)};
    \node(Ue3) at (16*4/56+2,-4) {$U_{e3}$};
     \path(16*4/56+1.1,-4)node{\footnotesize($0.06$)};
    \draw[->](MZ)--(Vcb); \draw[->](mt)--(Vcb); \draw[->](mu)--(Vcb);
    \draw[->](MZ)--(Vtd); \draw[->](mt)--(Vtd); \draw[->](mc)--(Vtd);
    \draw[->](MZ)--(Vub); \draw[double, ->](mt)--(Vub);
    \draw[->](ML)--(Um3); \draw[->](m3)--(Um3); \draw[->](m1)--(Um3);
    \draw[->](ML)--(Ut1); \draw[->](m3)--(Ut1); \draw[->](m2)--(Ut1);
    \draw[->](ML)--(Ue3); \draw[double, ->](m3)--(Ue3);
\end{tikzpicture}
$$
\caption{\small\baselineskip=13pt
A schematic depiction of the iteratively generated mass-scales in Table~\ref{t:Masses}, their horizontal position indicating their magnitude on a logarithmic scale. Masses indicated by $^\star$ involve additional numerical parameters; see Table~\ref{t:Masses} and text. For the mixing elements,
 $\color{purple}V_{ij}$ and $\color{purple}U_{ij}$ separately, their relative horizontal positions indicate their $\log_{\sss(M_{CH}/M_P)}({\cdots})$ values.}
 \label{f:Masses}
\end{figure}
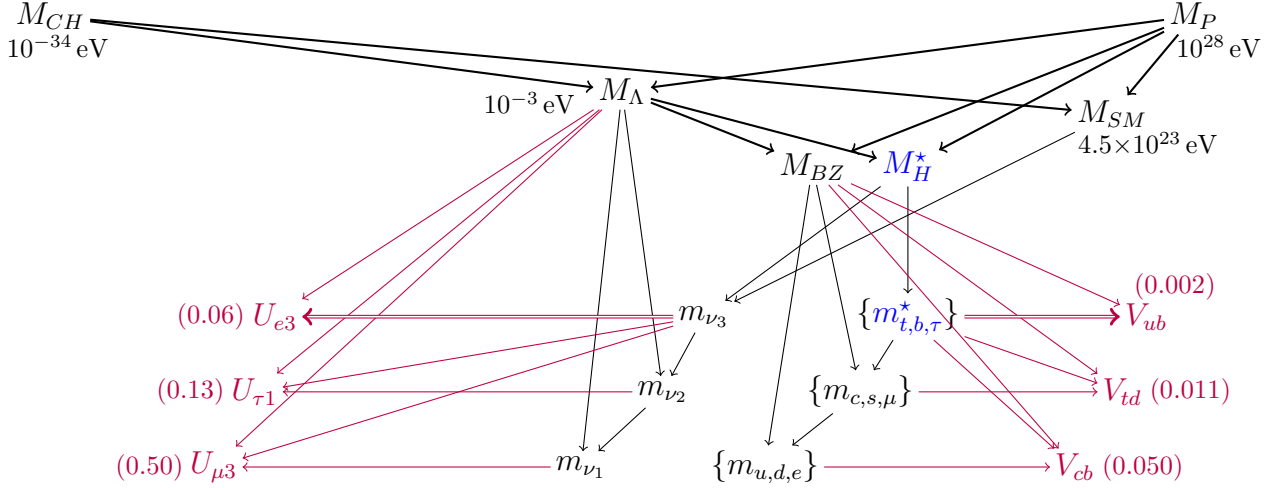

This hallmark UV/IR-mixing (which evidently transcends EFT),
and quantum contextuality (in contradistinction to anthropic reasoning\cite{Weinberg:1987dv}) are key features of our solution to the vacuum energy problem. It also involves the Born geometry\cite{Freidel:2013zga} (a kind of generalized 
mirror symmetry\cite{Berglund:2021hbo}) of the stringy (chiral phase-space-like) spacetime and its modular polarization, the Bekenstein\cite{Bekenstein:1980jp}, i.e., holographic\cite{tHooft:1993dmi,Susskind:1994vu,Fischler:1998st} bound, 
combined with stringy modular invariance
\cite{Polchinski:1998rq, Polchinski:1998rr}, 
and a stringy formula for the Higgs mass\cite{Abel:2021tyt}. 
These concepts are in fact crucial to addressing the problem 
of SM fermion masses (gravitational charges) and mixing matrices
in the context of string theory/quantum gravity, and lead to observations and mechanisms that are all realized within string theory --- the only consistent perturbative theory of both quantum gravity and SM-like matter. They
give rise to the concrete results summarized in Table~\ref{t:Masses} (depicted in Figure~\ref{f:Masses}), which
correlate very well with Bjorken's recent work\cite{Bjorken:2013aa,rBJ-MM}.
\begin{table}[h!tb]
\caption{\small\baselineskip=13pt 
The iterative estimation\cite{Berglund:2023gur} of the Standard Model mass hierarchy filigree, rewritten in terms of the cosmological horizon scale $M_{CH}$ all given in (3+1-dimensional) Planck units $M_P$. (Note that the Yukawa couplings in the table in principle originate from the criticality of the Standard Model as discussed in Section~\ref{s:PPP}.)
We thank Per Berglund for conversations on this topic.}
\setbox9\hbox{\tiny\cite{Abel:2021tyt}}
\setbox8\hbox{\footnotesize\cite{rPDG22}}
\setbox7\hbox{\footnotesize\cite{Bjorken:2013aa}}
\setbox6\hbox{\footnotesize\cite{Esteban:2020aa}}
\setbox5\hbox{\footnotesize\cite{Berglund:2023gur}}
\vspace*{.5\baselineskip}
\small\centering
$
\begin{array}{@{}r@{~}l@{~}l@{~}l@{~}l@{}}
 \multicolumn2l{\textbf{Observable}} 
 &\textbf{Reduced Form} &\textbf{Estimate} &\textbf{Value}\copy8 \\[-2pt]
\toprule\noalign{\vglue-2pt}
M_\Lambda
 &\simeq\sqrt{M\,M_P}
 &=M_{CH}^{1/2}\,M_P^{1/2}
 &\sim1.1{\times}10^{-3}\,\eV \\
M_H
 &\overset{\!\!\copy9}{=}
        \2{\xX}\, \sqrt{M_\Lambda\,M_P}
 &\simeq M_{CH}^{1/4}\,M_P^{3/4}\,\2{\xX}
 &\sim125\,\GeV \\
M_{BZ}
 &\sim \sqrt[3]{(M_\Lambda/2\pi)^2\,M_P}
 &\simeq M_{CH}^{1/3}\,M_P^{2/3}\,(2\pi)^{-\frac23}
 &\sim7.2\,\MeV\\
M_{SM}
 &\text{(see text)}
 &\define M_{CH}^{1/14} M_P^{13/14}
 &\sim4.5{\times}10^{14}\,\GeV\\
\midrule\noalign{\vglue-2pt}
m_t
 &=\2{Y_t}\,M_H
 &\simeq M_{CH}^{1/4}\,M_P^{3/4}\,\2{Y_t\,\xX}
 &\sim175\,\GeV &(173\,\GeV)\\
m_c
 &\sim\sqrt{M_{BZ}\,\2{m_t}}
 &\simeq M_{CH}^{7/24}\,M_P^{17/24}\sqrt{\2{Y_t\,\xX}}/(2\pi)^{1/3}
 &\sim1.13\,\GeV &(1.27\,\GeV) \\
m_u
 &\sim M_{BZ}^2/m_c
 &\simeq M_{CH}^{3/8}\,M_P^{5/8}\big/2\pi\sqrt{\2{Y_t\,\xX}}
 &\sim46.4\,\keV &(2.16\,\MeV)\\
m_b
 &=\2{Y_b}\,M_H
 &\simeq M_{CH}^{1/4}\,M_P^{3/4}\,\2{Y_b\,\xX}
 &\sim4.18\,\GeV &(4.18\,\GeV) \\
m_s
 &\sim\sqrt{M_{BZ}\,m_b}
 &\simeq M_{CH}^{7/24}\,M_P^{17/24}\sqrt{\2{Y_b\,\xX}}/(2\pi)^{1/3}
 &\sim174\,\MeV  &(93.4\,\MeV)\\
m_d
 &\sim M_{BZ}^2/m_s
 &\simeq M_{CH}^{3/8}\,M_P^{5/8}\big/2\pi\sqrt{\2{Y_b\,\xX}}
 &\sim301\,\keV &(4.67\,\MeV)\\
m_\tau
 &=\2{Y_\tau}\,M_H
 &\simeq M_{CH}^{1/4}\,M_P^{3/4}\,\2{Y_\tau\,\xX}
 &\sim1.75\,\GeV &(1.78\,\GeV) \\
m_\mu
 &\sim\sqrt{M_{BZ}\,m_\tau}
 &\simeq M_{CH}^{7/24}\,M_P^{17/24}\sqrt{\2{Y_\tau\,\xX}}/(2\pi)^{1/3}
 &\sim113\,\MeV &(106\,\MeV) \\
m_e
 &\sim M_{BZ}^2/m_\mu
 &\simeq M_{CH}^{3/8}\,M_P^{5/8}\big/2\pi\sqrt{\2{Y_\tau\,\xX}}
 &\sim464\,\keV &(511\,\keV)\\
\midrule\noalign{\vglue-2pt}
m_{\nu3}
 &\sim M_H^2/M_{SM}
 &\simeq M_{CH}^{3/7}\,M_P^{4/7}\,\2{\xX^2}
 &\sim3.5{\times}10^{-2}\,\eV & (<.8\,\eV)\\
m_{\nu2}
 &\sim \sqrt{M_\Lambda\,m_3}
 &\simeq M_{CH}^{13/28}\,M_P^{15/28}\,\2{\xX}
 &\sim6.2{\times}10^{-3}\,\eV & (<.8\,\eV)\\
m_{\nu1}
 &\sim M_\Lambda^2/m_2
 &\simeq M_{CH}^{15/28}\,M_P^{13/28}\big/\2{\xX}
 &\sim2.0{\times}10^{-4}\,\eV & (<.8\,\eV)\\
\midrule\noalign{\vglue-2pt}
|V_{cb}|
 &\sim\frac{M_{BZ}}{\sqrt{m_b m_d}}\approx0.050^\dag
 &\simeq(M_{CH}/M_P)^{1/48}\big/(2\pi)^{1/6}(\2{Y_b\,\xX})^{1/4}
 &\sim 0.204 &(0.041)\copy7\\
|V_{td}|
 &\sim\frac{M_{BZ}}{\sqrt{m_b m_s}}\approx0.011^\dag
 &\simeq(M_{CH}/M_P)^{3/48}\big/\sqrt{2\pi}(\2{Y_b\,\xX})^{3/4}
 &\sim 0.008 &(0.008)\copy7\\
|V_{ub}|
 &\sim\frac{M_{BZ}}{\sqrt{m_b m_b}}\approx0.002^\dag
 &\simeq(M_{CH}/M_P)^{4/48}\big/(2\pi)^{2/3}(\2{Y_b\,\xX})
 &\sim 0.002 &(0.003)\copy7\\
%\midrule[.2pt]
|U_{\mu3}|
 &\sim\frac{M_\Lambda}{\sqrt{m_3 m_1}}\approx0.50^\dag
 &\simeq(M_{CH}/M_P)^{1/56}\big/\2{\xX}^{1/2}
 &\sim 0.422 &(0.63)\copy6\\
|U_{\tau1}|
 &\sim\frac{M_\Lambda}{\sqrt{m_3 m_2}}\approx0.13^\dag
 &\simeq(M_{CH}/M_P)^{3/56}\big/\2{\xX}^{3/2}
 &\sim 0.075 &(0.26)\copy6\\
|U_{e3}|
 &\sim\frac{M_\Lambda}{\sqrt{m_3 m_3}}\approx0.06^\dag
 &\simeq(M_{CH}/M_P)^{4/56}\big/\2{\xX}^2
 &\sim 0.032 &(0.14)\copy6\\
\bottomrule
\multicolumn5l{\text{\color{blue}\footnotesize
$^\star$\,$\xX\define\sqrt{\tfrac{|\vev\cX|}{8\p^2}}\!\approx3.41{\times}10^{-2}$\cite{Abel:2021tyt},~ 
 $Y_t\approx7/5$,~
 $Y_b\approx1/30=Y_t/42$~ and~
 $Y_\tau\approx1/70\approx Y_t/100$}}\\ 
\multicolumn5l{\text{\footnotesize
$^\dag$\,These evaluations use\copy8-quark masses and a slightly higher estimate $M_{SM}\sim10^{15}\,\GeV$\copy5.}}
\end{array}
$
\label{t:Masses}
\end{table}
Our analysis of their correlations is what underlies a formulation of what quantum gravity ought to be --- a {\em gravitized quantum theory,}
as announced in the title.

This review builds on our previous paper 
\cite{Berglund:2023gur} (especially \SS\SS\,\ref{s:CC} and~\ref{s:PPP}) and is organized as follows:
Key aspects of the cosmological constant (cc) problem are discussed in \SS\,\ref{s:CC}, which we review (\SS\,\ref{s:CCP}) first in point-particle quantum field theory (QFT), and then also in string theory.
Its recent resolution\cite{Freidel:2022ryr} (see also\cite{Berglund:2022qsb}) is then presented in \SS\,\ref{s:CC-fix},
exhibiting that it
 ({\small\bf A})~connects quantized phase space properties with the Bekenstein (holographic) bound in a gravitational setting\cite{tHooft:1993dmi}, and
 ({\small\bf B})~is realized
 both in QFT (\SS\,\ref{s:QFT2}),
 as well as in string theory (\SS\,\ref{s:mStr}).
Given the computation of the vacuum energy based on (in general) dynamical phase space, we address the general argument for the formulation of quantum gravity as gravitized quantum theory in \SS\,\ref{s:QG=GQ}. Here we usher the reader from the concepts of modular spacetime as a quantum model of spacetime and the associated Born geometry to the metastring formulation of the general non-commutative and T-duality covariant string theory in dynamical quantum spacetime and dynamical Born geometry. Viewing quantum gravity as gravitized quantum theory also sheds light on the origins of quantum field theory and quantum mechanics, both based on the Born rule (Bornian quantum theory).
Distinguishing experimental probes of this new viewpoint are outlined in \SS\,\ref{s:EspGQ}: The most important ``smoking gun'' experimental probe of gravitized quantum theory is via higher order quantum interference (non-Born-rule-based) phenomena (non-Bornian quantum theory). We also discuss how gravitational interferometry can probe the statistics of spacetime quanta.
Returning to the SM, we discuss the recently obtained seesaw formula for its hallmark mass-scale, the Higgs mass, and in turn its intricate fermion mass and mixing structure in \SS\,\ref{s:PPP}:
A cascade of analogous seesaw formulae (see Figure~\ref{f:Masses} here and \SS\,\ref{s:Generalia} below) generates the entire SM-fermion mass hierarchy (\SS\,\ref{s:fMasses}) and fermion mixing (CKM and PMNS) angles (\SS\,\ref{s:fMixing}).
These results are thus shown to impose correlating bounds on all mass-scales, the cosmological constant ($\Lcc=(4\pi M_{CH}/\hbar c)^2$), the Higgs mass ($M_H$) as well as the masses and mixing of quarks and leptons. 
Furthermore, all these results should be understood as explicit results of quantum gravity viewed as generalized (gravitized) quantum theory.
Finally, \SS\,\ref{s:Coda} collects our concluding comments.

\section{From Vacuum Energy to Quantum Gravity}
\label{s:CC}

\subsection{The Cosmological Constant Problem}
\label{s:CCP}
We find that an explicit understanding of the vacuum energy (cosmological constant)
problem opens a new view on the problem of quantum gravity that also leads to novel empirical probes.
So, let us start by discussing the canonical
calculations of the cosmological constant in quantum field theory and in string theory by emphasizing various similarities and indicating the key differences between these two calculations. We follow the presentation from our 
recent paper\cite{Berglund:2023gur}.

\paragraph{Quantum Field Theory (QFT):}
The vacuum partition function of a free scalar in $D$ 
spacetime dimensions (which can be generalized for other fields) is
\begin{equation}
    Z_{\text{vac}} 
    = \int\rD\phi~ e^{-\int \frac{1}{2}\phi (-\partial^2 + m^2) \phi} 
    \propto \frac1{\sqrt{\det(-\partial^2 + m^2)}},
\end{equation}
and this expression can be rewritten as
\begin{equation}
    Z_{\text{vac}} =  e^{-\frac{1}{2}{\rm Tr\, log}(-\partial^2 + m^2)}.
\end{equation}
Fourier transforming to momentum space produces $-\partial^2 = k^2$, and
so\cite{rGSW2,Polchinski:1998rq}
\begin{equation}
    -\frac{1}{2} \log(k^2+m^2) = \int \frac{\rd \tau}{2 \tau} e^{-(k^2+m^2)\tau/2}.
\end{equation}
Here, the Schwinger parameter, $\tau$, is a worldline affine parameter (world-line time) of the particle, i.e., the quantum of the field $\phi$.
Tracing then produces
\begin{equation}
    \int \frac{\rd^D k}{(2\pi)^D}\log(k^2 + m^2) 
    = \int \frac{\rd^{D-1}k}{(2\pi)^{D-1}} \frac{\omega_k}{2},
\end{equation}
because
\begin{equation}
    \int \frac{\rd \tau}{2 \tau}\int \frac{\rd k^0}{2\pi}e^{-(k^2+m^2) \tau/2} = \frac{\omega_k}{2}
    \qquad\text{with}\qquad 
    \omega_k^2 \define k^2+m^2,
\end{equation}
where $\omega_k$ is equivalent to $k_0$ on-shell.
Thereby, we arrive at the familiar result for the vacuum energy density in $D$ spacetime dimensions 
\begin{equation}
    \rho_0 = \int \frac{\rd^{D-1}k}{(2\pi)^{D-1}} \frac{\omega_k}{2} \sim \Lambda_D,
\end{equation}
identifying $\Lambda_D$ as the {\em momentum space} volume.
 This inherently divergent expression leads to the infamous cosmological constant problem (see also \cite{Donoghue:2020hoh}).\footnote{In $D=4$ dimensions, Einstein's equations imply the cosmological constant to be $\Lambda_{cc}\sim\rho_0 G_N \sim \rho_0 l_P^2$.}
However, as emphasized by Polchinski, the vacuum partition function is also given as\cite{Polchinski:1985zf,Polchinski:1998rq}
\begin{equation}
        Z_{\text{vac}} =\bra{0}|e^{-iH t}|{0}\ket = e^{-i\rho_0 V_D},
\end{equation}
with $\rho_0$ the vacuum energy density and $V_D$ the $D$-dimensional {\em spacetime} volume.
On the other hand, $Z_{\text{vac}}=\exp\{Z_{S^1}\}$, where $Z_{S^1}$ is the circle ($S^1$) partition function in the world-line (particle) formulation:
\begin{equation}
    Z_{S^1}\define \int \frac{\rd \tau}{2 \tau}~ Z_{S^1}(\tau),\qquad
    Z_{S^1}(\tau)
     = V_D \int\frac{\rd^D k}{(2\pi)^D}~ e^{-(k^2+m^2)\frac{\tau}{2}}.
 \label{e:Z1}
\end{equation}
Combining the two expressions, the vacuum energy density becomes
\begin{equation}
        \rho_0 = \frac{iZ_{S^1}}{V_D} \sim \Lambda_D.
\end{equation}
This is an important formula that we will crucially utilize in what follows.

\paragraph{String Theory:}
Moving on to the case of a bosonic string (with obvious fermionic generalizations), we can use the result for the particle vacuum energy, albeit now to an infinite tower of particles with the stringy mass spectrum \cite{Polchinski:1985zf,Polchinski:1998rq}
\begin{equation}
        m^2 = \frac{2}{\alpha'}(h+\bar h - 2).
\end{equation}
Summing over the physical string states (``phys.\,st.'') then yields
\begin{equation}
        \sum_{\rm phys.\,st.} Z_{S^1} = \sum_{h,\Bar{h}} V_D \int \frac{\rd r (2\pi r)^{-D/2}}{2 r}\int \frac{\rd \theta}{2\pi}e^{i(h - \bar h)\theta} e^{-\frac{2}{\alpha'}(h+\bar h -2)\frac{r}{2}},
\end{equation}
with level-matching ($h=\bar{h}$, i.e., $\delta_{h,\bar{h}}$) imposed after the $k$-integration.
Defining $\tau\equiv\tau_1 + i \tau_2\define\theta + i \frac{r}{\alpha'}$ and $q\define e^{2\pi i\tau}$ as usual, the partition function of a bosonic string on $T^2$ (which can
be derived directly from the Polyakov path integral\cite{Polchinski:1985zf,Polchinski:1998rq}) becomes
\begin{equation}
        Z_{T^2} = V_D \int \frac{\rd \tau \rd \bar{\tau}}{2\tau_2} (4\pi^2 \alpha' \tau_2)^{-D/2} \sum_h q^{h-1} \bar{q}^{\bar{h}-1}.
\end{equation}
Let $r\define\alpha'\tau_2$, so that
\begin{equation}
        (4\pi^2\alpha'\tau_2)^{-D/2} =\int\frac{\rd^Dk}{(2\pi)^D}e^{-k^2\frac{r}{2}}.
\end{equation}
Akin to the case of particles or QFT, this produces:
\begin{subequations}
 \label{e:Z2}
\begin{equation}
    Z_{T^2} \define V_D \int\frac{\rd^D k}{(2\pi)^D} f(k^2) \define \int \frac{\rd \tau \rd \bar{\tau}}{2\tau_2} Z_{T^2} (\tau)\sim V_D\, \Lambda_D,
\end{equation}
with 
\begin{equation}
    \Lambda_D \define \int\frac{\rd^D k}{(2\pi)^D}
    \quad\text{and}\quad 
    f(k^2) \define \int_F \frac{\rd^2\tau}{2\tau_2} e^{-k^2 \alpha' \tau_2/2} \sum_h q^{h-1} \bar{q}^{\bar{h}-1},
\end{equation}
\end{subequations}
where $F$ is the fundamental domain. 
Since $f(k^2)$ is dimensionless, it cannot change the scaling of $Z_{T^2}$, so the vacuum energy is now $\rho_0 \sim Z_{T^2}/V_D$.
The crucial point is that the relevant partition function ($Z_{T^2}$ for strings, $Z_{S^1}$ for particles) scales as the {\it phase space} volume {\em both} in string theory and in QFT.
The key (and just as crucial) {\em difference} is in the region of integration:
\begin{equation}
  \text{QFT:}~~|\tau_1|<\tfrac{1}{2}~\&~\tau_2>0
  \qquad\text{vs.}\qquad
  |\tau_1|<\tfrac{1}{2}~\&~|\tau|>1 ~~\text{in string theory},
\end{equation}
where the difference, $\t_2>0 \to |\tau|>1$, is forced by modular invariance of string theory. Hence string theory (and thus quantum gravity) naturally cuts itself off at short distances.

This key difference renders the cosmological constant {\em UV-finite\/} in string theory (which makes physical sense, given the extended nature of the string), but is still related to $\rho_0\sim Z_{T^2}/V_D\sim \Lambda_D$, so that the ``cosmological constant problem'' persists in a manner very similar to what we have already encountered for particles or QFT.
Recall that supersymmetry (SuSy) cannot help with this fundamental problem:
Unbroken SuSy is well known to imply flat/Minkowski or anti~de~Sitter ($\L_D<0$) spacetime, due to the cancellation of the bosonic and fermionic contributions (and by including fluxes, in the case of anti~de~Sitter space), without changing the offensive (and generically large) phase space term in the above expression for the vacuum energy. The cosmological constant may be rendered positive only via SuSy breaking, which however does not change the relationship $\rho_0 \sim \Lambda_D$.
The crucial spacetime and momentum space volume factors, which are the ultimate cause of the cosmological constant problem, are not affected by SuSy, whether broken or not.

\subsection{Resolving the Cosmological Constant Problem}
\label{s:CC-fix}
We now review a new approach for addressing the cosmological constant problem using the above insights on the vacuum energy and spacetime and momentum space volumes.
 Following\cite{Freidel:2022ryr, Berglund:2023gur} (see also\cite{Berglund:2022qsb, Freidel:2023ytq}), we reconsider~\eqref{e:Z1} defining $Z_{S^1}$, and set $m=0$ for simplicity, and denote the momentum $p$.
Our discussion in fact holds also for all $m \neq 0$, other fields (not just scalars), as well as for the one-loop string partition function, $Z_{T^2}$.
 Therefore, let us consider
\begin{equation}
    Z_{S^1} = V_D \int \frac{\rd \tau}{2\tau}
    \int\frac{\rd^Dp}{(2\pi)^D}~e^{-\frac{p^2 \tau}{2}}.
\end{equation}
The spacetime volume being $V_D = \int \rd^D q$ naturally implies the {\em phase space} rewriting:
\begin{equation}
     Z_{S^1} = \int\frac{\rd \tau}{2\tau}~ Z(\tau),\quad
    Z(\tau) = \iint\frac{\rd^D q\;\rd^D p}{(2\pi)^D}~ e^{-\frac{p^2\tau}{2}} \,\define\, \Tr\big[e^{-\frac{p^2\tau}{2}}\big],
\label{e:Z1ph}
\end{equation}
where the $\tau$-integration is deferred to the very last step in the calculation, and $\Tr$ is now clearly defined over the {\em phase space}.

\paragraph{Modular Regularization of Phase Space:}
We proceed as follows to regularize the above phase space expression,
\begin{equation}
    Z(\tau) = \prod_{i=1}^4 \iint_{-\infty}^\infty
    \frac{\rd q_i\; \rd p_i}{2\pi}~
    e^{-\frac{p_i^2\tau}{2}},
 \label{e:Z(t)}
\end{equation}
in the case of four-dimensional ($D=4$) spacetime and momentum space.
Discretizing phase space and writing $\tx\define p/\epsilon$ and $x\define q/\lambda$, where the two scales are related by\,\footnote{\label{fn:chbar}Writing out $\hbar$ explicitly, we emphasize that $\epsilon$ and $\lambda$ are, respectively, {\em momentum\/}- and {\em length\/}-scales.} $\lambda\,\epsilon= \hbar$, this becomes
\begin{equation}
    Z(\tau) = \bigg( \lambda \,\epsilon \sum_{k,\tilde{k}\in {\mathbb{Z}}} \iint_0^1
    \frac{\rd  x\,\rd \tilde{x}}{2\pi}~ e^{-\frac{(k+\tilde{x})^2\epsilon^2\tau}{2}} \bigg)^4.
 \label{e:Z(t)2}
\end{equation}
Although this is still divergent, it admits
{\em modular regularization}\cite{Freidel:2015pka,Freidel:2015uug}
by restricting the sum to a finite range\footnote{\label{fn:horizons}This restriction is motivated also on physical measurement/observable grounds: spacetime integration is limited by the finite cosmological horizon, while momenta are limited by the Planck scale since probes of higher momentum are self-consistently invisible in the Planckian region around the event horizon of the interacting probe-target system.}:
\begin{equation}
    Z(\tau) = \left(\lambda \,\epsilon
    \sum_{k=0}^{N_q-1} \sum_{\tilde{k}=0}^{N_p-1} \iint_0^1 \frac{\rd x\,\rd \tilde{x}}{2\pi}~
    e^{-\frac{(k+\tilde{x})^2\epsilon^2\tau}{2}} \right)^4.
 \label{e:MRFR}
\end{equation}
This identifies $N_q,N_p$ as counting, respectively, the number of spacetime and momentum space ``unit cells,''\footnote{From the modular polarization\cite{Freidel:2016pls} point of view, these count {\em unit cells of vacua,} not on-shell states or particles.}
and prompts defining:
\begin{equation}
     l\define N_q \lambda,\quad \text{and}\quad
     M_\L \define N_p \epsilon,
     \qquad\text{with}\quad N=(N_p N_q)^4\in \mathbb{Z}.
 \label{e:N}
\end{equation}
Thereby, $l^4\define V_4$ is the size (4-volume) of spacetime,
and $M_\L^4$ is the size (4-volume) of momentum space.
Furthermore (see footnote~\ref{fn:chbar}),
\begin{equation}
    l^4 M_\L^4 = N\,\hbar^4,
    \quad \text{or}\quad
    M_\L^4 = \frac{N\,\hbar^4}{l^4}.
 \label{e:lLN}
\end{equation}
However, $e^{-p^2\tau/2}\leq 1$ in~\eqref{e:Z(t)} implies an {\em upper bound}: $\rho_0\sim M_\L^4\leq \frac{N}{l^4}$, for $D=4$.
Our above calculation of the partition function of the bosonic string on $T^2$ in $D=4$ implies that the same bound also holds in string theory:
\begin{equation}
    \rho_0 \leq \frac{N}{l^4} .
 \label{e:rho-bound}
\end{equation}
Stemming from the phase-space modular regularization~\eqref{e:MRFR} and as we will discuss further in the following subsection,
this result extends to quantum fields and effective potentials in QFT, including the cosmological phase transitions (electroweak and QCD), without changing the outcome for this bound on the vacuum energy.

\paragraph{Holography:}
Consider now the Bekenstein bound in a four-dimensional space with a cosmological horizon with the above positive cosmological constant, i.e., in de~Sitter spacetime.
This spacetime metric is in static coordinates:
\begin{equation}
    \rd s^2_{dS} = - \Big(1-\frac{r^2}{r_{CH}^2}\Big) \rd t^2
    + \frac{\rd r^2}{\big(1-\frac{r^2}{r_{CH}^2}\big)}+r^2\rd \omega^2_{S^2},
\end{equation}
where the cosmological horizon, $l\define r_{CH}$, is the size of the observed spacetime.
Following the discussion in\cite{Freidel:2022ryr,Freidel:2023ytq},
identifying the above microscopic counting of ground states with the gravitational entropy renders the Bekenstein bound  ($S_{\text{grav}} = l_P^{-2}\textit{Area}$) as\cite{Bekenstein:1980jp}\footnote{In our previous work on string theory in de Sitter space (section 4.6 of\cite{Berglund:2022qsb}) we have pointed out that modular quantization and the appearance of the new quantum number $N$ imply the holographic scaling, and thus this IR condition is fundamentally tied to the nature of modular polarization.}:
\begin{equation}
    N\leq \frac{l^2}{l_P^2}.
 \label{e:NllP}
\end{equation}
Combining this holographic bound with the phase space bound~\eqref{e:rho-bound} on $\rho_0$ leads to
\begin{equation}
    \rho_0\leq \frac{1}{l^2\, l_P^2},
  \label{e:rho=llP}
\end{equation}
which exhibits a mixing of the UV ($l_P$) and the IR ($l$) scales.
This mixing between the short distance and long distance scales induces a {\em bound} for the cosmological constant, which in $D=4$ dimensions ($\Lambda_{cc}=\rho_0\, l_P^2$) reads:
\begin{equation}
    \Lambda_{cc} \leq \frac{1}{l^2}.
\end{equation}
Associated with the vacuum energy density is a natural energy scale, $\epsilon_{cc}$:
\begin{equation}
    \rho_0 = \epsilon_{cc}^4\sim \frac{1}{l^2\, l_P^2}.
\end{equation}
Its corresponding natural length scale, $l_{cc}\simeq 1/\epsilon_{cc}$, is given by the the seesaw formula
\begin{equation}
    l_{cc}\simeq \sqrt{l\, l_P},
    \qquad\text{i.e.,}\qquad
    M_\L\simeq\sqrt{M_{CH}\,M_P},
  \label{e:lcc=llP}
\end{equation}
where $M_{CH}\sim10^{-34}$\,eV is the Hubble (cosmological horizon) mass-scale and $M_P \sim 10^{19}$\,GeV, is the Planck scale.
Finally, note that the integration over the world-line or the world-sheet parameters does not change this final result. These integrations done naturally at the end only provide an overall renormalization of the Newton constant.

Extending from\cite{Berglund:2023gur}, we summarize the remarkable properties of the above results:
\begin{itemize}\itemsep=-3pt\vspace*{-1mm}
  \item Using $l\sim 10^{27\text{--}28}$\,m (the observed cosmological horizon) and $l_P \sim 10^{-35}$\,m (as a fundamental length unit), the seesaw relation~\eqref{e:lcc=llP} yields $l_{cc}\simeq 10^{-4}$\,m
  (i.e., $\epsilon_{cc}\simeq 10^{-3}$\,eV),
  in agreement with observations and identifying $\epsilon_{cc}=M_\L$ from~\eqref{e:N}.
  \item In the $l\to\infty$ limit, \eqref{e:rho=llP} forces $\rho_0\to 0$, and $l$ is the IR length-scale. Conversely, $\rho_0$ (and $\Lcc$) is nonzero and small because the cosmological horizon is finite and large.
  \item The relation~\eqref{e:lcc=llP} is radiatively stable since it is UV-independent; $l_P$ here serves as a reference length unit.
  \item The extraordinary smallness of the cosmological constant thereby owes, essentially, to the universe admitting a large number of degrees of freedom: $N \sim 10^{124}$. The entropic/area nature of the Bekenstein bound,~\eqref{e:NllP}, relates $N$ to the {\em square} of the length-ratio, $l/l_P\sim10^{61\text{--}62}$, although $N$ counts the number of degrees of freedom in the 4-dimensional spacetime volume.
  \item In turn, the number of degrees of freedom in the universe ($N$) is large as that makes scale of fluctuations small, $\frac{1}{\sqrt{N}}$, indicating its stability.
  \item This estimates $N_i\sim N^{1/4} \sim 10^{31}$ (where $i$ is $t,x,y,z$). Not so unreasonable in comparison with Avogadro's number ($10^{23}$) for matter degrees of freedom, it is tempting to refer to $N_i\sim10^{31}$ as the spacetime Avogadro number. This should be measurable in the context of gravitational interferometry (see Section~\ref{s:EspGQ}). From this point of view, the characteristic scale $l_{cc}=N_i\,l_P\simeq10^{-4}$\,m associated with the vacuum energy (and also characteristic of certain extra dimension models) is the collective, macroscopic spacetime scale, tied to the spacetime Avogadro number of $10^{31}$.
\end{itemize}
Furthermore, we emphasize the quantum contextuality of the above calculation: The measurement of a quantum observable depends on which commuting set of observables are within the same measurement set of observables, 
i.e., quantum measurements depend on the {\em context\/} of measurement. The concept of contextuality will be crucial in our analysis of masses and mixing angles of elementary particles in Section~\ref{s:PPP}.
\begin{itemize}\itemsep=-3pt\vspace*{-1mm}
    \item First, the momentum-scale $\epsilon$ is {\em not} a cut-off, since $\epsilon$ and $\lambda$ can be arbitrary, albeit reciprocally related by $\lambda \,\epsilon =\hbar$; see footnote~\ref{fn:chbar} on p.~\pageref{fn:chbar}.

    \item Second, $\epsilon^4$ is effectively eliminated in favor of $N$, which is the new quantum number, and the size of spacetime, $l=r_{CH}$, the cosmological horizon, i.e., the size of the observed classical spacetime.

    \item In turn, $N$ is determined by the Bekenstein bound,~\eqref{e:NllP}, and is thereby related to $l$ and $l_P$ (the ultimate IR and UV scales, respectively) --- which is where gravity enters, 
    via the familiar relation $G_N\sim l_P^{2}$.

\end{itemize}
In contradistinction, EFT cannot possibly ``see'' $N$, and in particular cannot ``know'' about either the Bekenstein bound or the UV/IR mixing.
For example, vacuum energy routinely cancels in the computation of EFT correlation functions. Also, EFT {\em is defined\/} in classical spacetime\cite{Weinberg:1974tw,Weinberg:1978kz}\footnote{The above seesaw relation  $l_{cc}\simeq \sqrt{l\, l_P}$ does appear in \cite{Cohen:1998zx},
where $l$ and $l_P$ are, respectively, the ultimate IR and UV length-scales in EFT, and are related by the physics of black holes/holographic bound. However, that approach has neither of the crucial aspects of our derivation: modular representation, the number of phase space cells $N$, the explicit UV/IR mixing and contextuality. We comment on the connection with EFT at the end of Section~\ref{s:QFT2}.}.
Thus the above calculation calls for a fundamental quantum formulation that relies on the modular polarization\cite{Freidel:2016pls}. Precisely this is provided by the metastring formulation of string theory\cite{Freidel:2015pka},
which we will duly discuss in Section~\ref{s:mStr}.

\subsection{The Vacuum Energy in QFT and Phase Space}
\label{s:QFT2}
The foregoing analysis and the cosmological constant bound~\eqref{e:lcc=llP} extend to QFT as it combines the Bekenstein bound and our phase space argument, both of which are universal and insensitive to any QFT/EFT vacuum redefinition such as due to possible phase transitions.

Relying on standard generalizations to other fields and even string theory, consider a scalar field theory\cite{rSW1, rAZ-QFT} with the typical Lagrangian
\be
L = \frac{1}{2} (\partial \phi)^2  - V(\phi)
\quad\text{with}\quad
V(\phi) = \frac{1}{2} m^2 \phi^2 + \frac{1}{24}g \phi^4
\ee
and its partition function
\be
Z(J) \define \int \rD \phi~ e^{i [S(\phi) + J\phi]}
\>\define e^{i W(J)}.
\ee
The generating functional of vacuum correlation functions, $W(J)$, is a direct analogue of~\eqref{e:Z1} and~\eqref{e:Z2}, defines the effective action via
the Legendre transform:
\be
\Gamma(\phi) \define W(J) - \int\rd^4x~ J(x) \phi(x)
~=\int\rd^4x\>[\,\ldots -V_{\text{eff}}(\phi) +\ldots\,],
\ee
which in turn defines the QFT vacuum energy as the minimum of $V_{\text{eff}}(\phi)$.
(This also absorbs the proper path integral normalization that is responsible for the vacuum energy.)

Expanding the original action $S(\phi)$ up to quadratic fluctuations and
Gaussian path-integration produces the corresponding $\hbar$-expansion of the effective action and potential,
\begin{align}
\Gamma(\phi) &= S(\phi) +\frac{i \hbar}{2}\,\Tr\log\big[\partial^2 +V''(\phi)\big]~ + O(\hbar^2),\\
V_{\text{eff}}(\phi) &= V(\phi) - \frac{i \hbar}{2}\int \frac{\rd^4 k}{(2 \pi)^4}\log[{k^2 -V''(\phi)}]~ + O(\hbar^2),
\end{align}
and to the famous Coleman-Weinberg potential\cite{Coleman:1973jx}
by explicit evaluation of the momentum integral.
With the Schwinger parametrization of the logarithm\cite{rGSW2,Polchinski:1998rq},
\be
\log\big[ U(k^2, \phi)\big]
= -\int \frac{\rd r}{r}~e^{-U(k^2, \phi)\,r/2},
\ee
quantum corrections to the effective action may again be rewritten, most crucially, as the phase space integral
\be
\Gamma(\phi)\sim \iint\frac{\rd^4 x\,\rd^4 k}{(2 \pi)^4}
\int\frac{\rd r}{r}~ e^{-U(k^2, \phi)\,r/2}
\quad\text{with}\quad
0<e^{-U(k^2, \phi)\,r/2}<1,
\ee
in full analogy
with~\eqref{e:Z1}--\eqref{e:Z2}--\eqref{e:Z1ph}.
Modular regularization~\eqref{e:MRFR} then again leads to the same bound on the vacuum energy evaluated from the effective action, upon coupling to gravity and being subject to the Bekenstein bound. Once again, the integration of the Schwinger parameter is done at the end, and its effect is absorbed in the renormalization of the Newton constant.

The canonical evaluation of the effective action, $\Gamma(\phi)$, at the effective potential minimum, implies a divergent vacuum energy the cosmological constant problem upon coupling to gravity. This evaluation is inherently sensitive to both radiative corrections and any phase transitions dictated by the effective potential. At finite temperature $T$, the familiar Landau-Ginsburg description (with $a,g>0$),
\be
L(\phi,T) = \frac{1}{2} (\partial \phi)^2  - V(\phi,T)
\quad\text{with}\quad
V(\phi,T) = \frac{1}{2} a(T-T_c) \phi^2 + \frac{1}{24}g \phi^4,
\ee
has a global minimum ($\phi=0$) above the critical temperature, $T_c$.
Below $T_c$, $L(\phi,T)$ develops new global minima, to one of which the system transitions from the now unstable (tachyonic) local maximum, $\phi=0$. Coupling the symmetry breaking order parameter, $\phi$, to a gauge field renders it massive in the classic Higgs mechanism.
The above analysis however still applies and again yields the quantum part of the effective action to scale as
\be
\Gamma (\phi, T) \sim \iint \frac{\rd^4 x\;\rd^4 k}{(2 \pi)^4}
\int \frac{\rd r}{r} e^{-U(k^2, \phi, T)\,r/2}
\quad\text{with}\quad
0<e^{-U(k^2, \phi, T)\,r/2}<1.
\ee
The above-established bound for the vacuum energy (determined by the minimum of this finite temperature effective potential) therefore continues to hold and gives the same seesaw formula~\eqref{e:lcc=llP} when combined with the Bekenstein bound and modular regularization. In particular, as standard in finite-temperature QFT, we replace
\be
\int (\,\cdots) \frac{\rd^4 k}{(2 \pi)^4} \to 
T \sum_{k^0=2 \pi i n T} \int (\,\cdots) \frac{\rd^3 k}{(2 \pi)^3},
 \label{e:k0->T}
\ee
adding to the effective potential:
\be
V_{\text{eff}} (T)\sim  \frac{T}{2}\sum_n \int\frac{d^3 k}{(2 \pi)^3}~
\log[\,{4\pi^2 n^2 T^2 +{\vec{k}}^2 +V''(\phi)}\,].
\ee
Using again the Schwinger parametrization produces
\be
\Gamma (\phi, T) \sim T  \sum_n\iint \frac{\rd^4 x\;\rd^3 k}{(2 \pi)^3}
\int \frac{\rd r}{r}~ e^{-[\,4\pi^2 n^2 T^2 + U({\vec{k}}^2, \phi, T)\,]\,r/2},
\ee
which is still bounded by the volume of phase space since $T$ measures the size of the ``imaginary time/energy'' direction and the indicated summation stems from having discretized that direction in~\eqref{e:k0->T}. As usual, returning to the continuum recovers the expected $\sum_n\int\rd^3k\to\int\rd^4k$. 

Throughout, the Schwinger parameter $r$-integration is left to the end of the calculation, where it merely renormalizes the Newton constant, and has no influence on the vacuum energy bound.
Before the $r$ integration, it is the phase space volume that bounds the effective action; together with the Bekenstein bound and modular regularization, this produces the bound on the cosmological constant as derived above.
The ubiquitous appearance of the phase space volume in these expressions for the effective action therefore justifies applying our argument from the previous section, which together with the Bekenstein bound and modular regularization reproduces the same seesaw formula~\eqref{e:lcc=llP} and cosmological constant bound, also in the context of QFT coupled to gravity.
These same characteristics of our bound imply its radiative stability and continued validity regardless of the cosmological phase transitions (electroweak, QCD). Conceptually (see also footnote~\ref{fn:horizons}), it is the mixing~\eqref{e:rho=llP} of the UV (gravitational, Planck scale) and the IR (cosmological horizon, Hubble scale) that insures this stability.
Neither local QFT nor any EFT can ``see'' either this UV/IR mixing or the ensuing resolution of the vacuum energy problem, since they by construction omit any global (non-local) features associated with modular regularization of phase space.

It is worth emphasizing that vacuum energy is in EFTs tied to the path integral normalization, and so cancels in usual EFT calculations (without gravity).
The standard EFT results are recovered in a constrained double scaling limit\cite{Berglund:2022qsb}, where 
$N,l \to \infty$ while $N/l^4\define 1/\lL^4= \textit{const.}\!<\infty$, and $l_P \to 0$.
This still preserves the seesaw nature of the formula for $\lL$, leading to the analogous observation of\cite{Cohen:1998zx}, where however $l$ and $l_P$ serve, respectively, as the EFT IR and UV cut-offs.
Unlike the crucial role of quantum spacetime degrees of freedom and the mixing between the UV and IR physics in quantum gravity, EFT is defined in classical (and not quantum) spacetime, and so is fundamentally insensitive to the UV/IR mixing.
Nevertheless, when combined with holography, EFT can capture the essential low energy features (such as the final geometric mean formula) of the fully quantum treatment of the vacuum energy problem, that is valid both in the UV and IR regions.

In conclusion of this section, the vacuum energy (cosmological constant) problem can be explicitly understood through a combination of phase space reasoning (with, in principle, a dynamical phase space) and the holographic bound (as an infrared requirement). 
This in turn opens a new vista on the theoretical and empirical foundations of quantum gravity.

\section{Quantum Gravity (QG) = Gravitized Quantum (GQ)}
\label{s:QG=GQ}
Following the new calculation of the vacuum energy that matches the observed value (interpreted as the cosmological constant in Einstein's gravitational equations), it becomes apparent that the essential new ingredient of quantum gravity is a dynamical quantum phase space, given the dynamical nature of spacetime as well as the fact that spacetime and momentum space appear on equal footing as slices of phase space. This suggests a dynamical form of the Born rule, implied by the geometry of quantum phase space, which should be used to define general quantum probabilities and observables of quantum gravity. The familiar Born rule should be likened to the Minkowski metric of special theory of relativity, which is maximally symmetric, homogeneous and isotropic, and is {\em a linearization\/} of the general dynamical spacetime geometry of general relativity that has no such restrictions. Thus, the main difference between traditional approaches to quantum gravity and the one presented in the present review, is the gravitization of quantum theory, that is a fully dynamical geometry and topology of the space of quantum states. In this section we present a general discussion of first, a non-dynamical quantum spacetime (called modular spacetime) which captures the geometry of quantum theory, and then extend this to a dynamical quantum spacetime formulation found in a non-commutative and phase-space like formulation of string theory (the metastring).

\subsection{Quantum Spacetime and Quantum Relativity}
\label{s:QSpT}
In this subsection we present a model of quantum spacetime, called modular spacetime based on the most general geometry of quantum theory, dubbed Born geometry. Born geometry (the geometry of quantum relativity, and a unification of symplectic, doubly orthogonal and doubly metric geometry) should be understood as a direct analog of Minkowski geometry in the context of classical relativity. As a matter of fact, we suggest the following analogy between
classical and quantum relativity.

{Classical relativity} can be understood as a logical progression from 
 {\em a\/})~special relativity --- motivated by classical field theory --- ({with Minkowski spacetime/geometry and relativity of simultaneity}) to
 {\em b\/})~relativistic field theory ({with, in the quantum context, unitary representations of the Lorentz/Poincar\'e symmetry and with the famous prediction of antiparticles and the spin-statistics relation}) and, finally, {\em c})~general relativity ({with dynamical classical spacetime}).
Here, spacetime relativity is the first (classical) relativity (with both spacetime and matter being classical). General relativity is a dynamical extension of the same.

Quantum relativity can be analogously understood as a logical progression from 
 {\em A\/})~Quantum mechanics (QM) understood from quantum spacetime (with {modular spacetime, Born geometry and relative (observer dependent) locality} to be explained in what follows) --- this is a Bornian quantum theory;
to 
 {\em B\/})~QFT (with {metafields} and the new prediction --- metaparticles --- this is an intrinsically non-commutative covariant formulation based on modular spacetime with both spacetime and dual spacetime as natural limits) --- this is also a Bornian quantum theory;
and, finally, 
 {\em C\/})~{ gravitized quantum theory --- a quantum analog of general relativity, a non-Bornian quantum theory} (with {dynamical quantum spacetime} and {dynamical Born geometry} as realized in the {metastring} formulation of an intrinsically non-commutative, T-duality covariant and (chiral) phase-space-like string theory with metaparticle zero modes. Metaparticles appear as natural dark matter quanta and the geometry of dual spacetime as the natural origin of dark energy. Similarly, general quantum statistics of spacetime quanta, as well as
 general higher order quantum correlations characterize gravitized quantum theory, which as a sort of metaquantum theory, also sheds light on the origin of QM and QFT.)
Here Quantum Mechanics (QM)/Quantum Field theory (QFT) can be viewed as second (quantum) relativity (with matter being quantum, and spacetime classical). 
In the same vein, Quantum Gravity (QG) = Gravitized Quantum theory (GQ) can be viewed as third (quantum gravity or gravitized quantum) relativity (with both spacetime and matter being quantum).
(``Third relativity'' is apparently the phrase advocated by Finkelstein and Wheeler; see David Finkelstein's book\cite{Finkelstein:2012Qua}.)

\subsection{Modular Variables and Polarization}
In the generic modular polarization (representation, picture) of quantum theory, instead of considering the standard commutation relations between the position and momentum operators, one considers the generators of translations in {\em phase space}
\begin{equation}
    \hat U_a = e^{\frac{i}{\hbar} \hat p \,a},\quad
     \hat V_{\frac{2\pi \hbar}{a}} = e^{\frac{i}{\hbar} \hat q \frac{2\pi \hbar}{a}}\quad  \implies
     [\hat U_a, \hat V_{\frac{2\pi \hbar}{a}}] =0.
\end{equation}
In terms of {\em modular variables} introduced by Aharonov and collaborators\cite{Aharonov:2005uc},
\begin{equation}
    [\hat q]_a \define \hat q\, {\rm mod}\, a\quad
    [\hat p]_\frac{2\pi \hbar}{a} \define \hat p\, {\rm mod}\, \frac{2\pi \hbar}{a}\quad \implies
    \big[\,[\hat q]_a, [\hat p]_\frac{2\pi \hbar}{a}\,\big]=0.
\end{equation}
The {\em space} of commuting subalgebras of the Heisenberg algebra, $[\hat q,\hat p]=i\hbar$, which in the covariant (self-dual lattice) phase space formulation becomes the modular spacetime\cite{Freidel:2015pka,Freidel:2015uug,Freidel:2016pls} {\em is\/} the target space of the metastring\cite{Freidel:2015pka} and 
the metaparticles \cite{Freidel:2017wst, Freidel:2017nhg, Freidel:2018apz}
to be explained in the next subsection.
For example, vertex operators in metastring theory are representations of this Heisenberg algebra. This description (intrinsically non-commutative, since $[x,\tilde x]=i$, where $x \equiv q/\lambda$ and $\tilde x \equiv p/\epsilon$ and $\lambda \epsilon = \hbar$) will appear in what follows in the metastring formulation of string theory which avoids all of the co-cycles that turn up in standard descriptions of 
the vertex operator algebra in string theory\cite{Polchinski:1998rq,Polchinski:1998rr}, the vertex operators being the above generators of translations in phase space.

A more elementary (and familiar) argument for the existence of modular spacetime may be presented as follows: In quantum theory, short (UV) distances are associated with high energy, as implied by the indeterminacy relation, $\delta q \sim 1/\delta p$ (in $\hbar=1$ units). On the other hand, in classical (as well as semiclassical) gravity, the Schwarzschild radius $R_S$ of a mass $M$ is given by $R_S \sim G M \sim l_P^2 M$,
where $G \sim l_P^2$ is the gravitational constant in 4-dimensional spacetime, with $l_P$ the Planck length.
In quantum gravity, quite generally, one therefore expects that higher energy leads to larger (IR) distances $\delta q\sim l_P^2\,\delta p$.
These diametrically contrasting behaviors (associated with UV and IR) may be reconciled by relating the UV and IR physics:
Recall that, given a fundamental lattice length, quantum states are described in terms of quantum numbers associated with both a lattice and its dual\cite{Freidel:2016pls}. In our present case, this involves momenta $p$ and their duals $\tilde{p}$, provided that these commute $[p, \tilde{p}]=0$.
The indeterminacies $\delta p$ and $\delta\tilde{p}$ thereby being interchangeable provides the first substitution in the chain:
 $l_P^2\,\delta p \sim \delta q \to l_P^2\,\delta \tilde{p} \sim \delta q
  \to l_P^2 (\delta \tilde{q})^{-1} \sim \delta q
  \Rightarrow \delta q\,\delta \tilde{q} \sim l_P^2$,
where the second replacement used the canonical
$\delta\tilde{p}\,\delta\tilde{q}\sim1$ indeterminacy relation.
This implies a new fundamental non-commutativity between spacetime and dual spacetime coordinates  $[q, \tilde{q}] \sim i l_P^2$.
The commutative nature of modular variables in quantum theory insures that this can be completely 
covariantized\cite{Freidel:2016pls}.
Thus, combining the fundamental quantum and gravitational relations between spatial distances and momenta  leads to:
\begin{itemize}
\itemsep=-3pt\vspace*{-1mm}
\item 
 the concept of dual spacetime,
\item 
 the fundamental non-commutativity between spacetime and dual spacetime,
 \item 
 the Heisenberg algebras:
$[q, \tilde{q}] = i l_P^2$,
$[q, p] = i$,
$[\tilde{q}, \tilde{p}] = i$,
$[p, \tilde{p}] =0$.
\end{itemize}

Note that {modular variables are covariant} (there is modular energy and modular time as well).
Take the fundamental length $\lambda$ and energy $\epsilon$, so that $\lambda \epsilon \equiv \hbar$.
{ Modular variables are non-local (but consistent with causality --- this is the origin of the uncertainty principle)}.
{\it We emphasize the notion of contextuality}: {\it in a double slit experiment the parameters $\lambda$ and $\epsilon$ are {contextual} to
the experiment}. This contextuality was an important point in our discussion of the cosmological constant problem and it will be important in Section~\ref{s:PPP} when we discuss masses and mixing angles of quarks and leptons.
{Also we stress the explicit non-locality}:  Take $H = \frac{p^2}{2m} + V(q)$ and write the Heisenberg equation of motion for $e^{ipR/\hbar}$, or equivalently $[p]_R$.
(Here $R$ is a contextuality parameter, such as the distance between two slits in the double slit interference experiment.)
\begin{equation}
    \frac{d [p]_R}{d t} = \cdots \frac{V(q +R/2) - V(q-R/2)}{R}
\end{equation}
Thus quantum mechanics can be understood to arise from consistency between non-locality (owing to modular variables, but ultimately having origin in quantum gravity viewed as gravitized quantum theory, or a metaquantum theory) and causality (that is, compatibility with the Lorentz symmetry).

\subsection{Modular Spacetime, Born Geometry and Quantum Physics}
Now, let us reformulate quantum mechanics (QM) using (covariant) modular variables via modular spacetime.
({Quantum theory tells us something new about quantum spacetime via the concept of modular spacetime.})
What precisely is {modular space}? 
{Modular space is the space of all commuting subalgebras of the Heisenberg-Weyl algebra.}
By definition $[q, p] =i\hbar$ is the Heiseinberg-Weyl algebra, whereas $[[q]_a, [p]_{2 \pi \hbar/a}]=0$ is the commuting subalgebra of Weyl-Heisenberg. Here we have the following fundamental result encapsulated in
{\it Mackey's Theorem}: {\it the space of all commuting subalgebras of the Heisenberg-Weyl algebra is a self-dual phase space lattice lifted to the Heisenberg-Weyl algebra.} This theorem allows us to define modular spacetime and the associated Born geometry.

In particular, if we use covariant modular variables we obtain {modular spacetime} of $d$ spacetime dimensions.
{In the above theorem the concept of phase space comes with the natural symplectic structure} $Sp(2d)$, $\omega_{ab}$.
{The concept of a self-dual lattice} ($\ell\oplus\tilde{\ell}$), where $\ell$ is a lattice implies {\it doubly-orthogonal} $O(d,d)$, $\eta_{ab}$.
Finally, to {define the vacuum} on this self-dual lattice, {we need doubly metric structure} $O(2, 2d-2)$, $H_{ab}$. The triple
$(\omega, \eta, H)$ {\it defines Born geometry.}
{Their triple intersection gives the Lorentz group.}
Thus {QM follows from non-locality (fundamental length/time of quantum gravity) that is consistent with causality (implied by the Lorentz symmetry).}
We emphasize that one can be localized (and thus local QFT is possible) in a particular phase space cell, but one can not tell in which phase space cell (this is the origin of the uncertainty principle),
because the number operators (for the spatial and the momentum directions, respectively) do not commute with modular variables.

{How can fundamental length/time be consistent with the Lorentz symmetry?} (In some sense, this is one of the main puzzles of quantum gravity.)
This is possible because of {\it relative (observer dependent) locality} \cite{Amelino-Camelia:2011lvm}). Different observers see different spacetimes (slices of modular spacetime).
{However, different spacetimes are in linear superposition, and so fundamental length/time is consistent with the Lorentz symmetry.}
(This is similar to spin: the superposition of up and down spin gives the Bloch sphere which is consistent with rotation symmetry, even though spin is discrete 
\cite{Freidel:2016pls}.) Thus linear quantum superpositions are needed to 
reconcile the fundamental length/time with the Lorentz symmetry. This is the quantum spacetime origin of quantum theory.

Next we introduce the generic quantum polarization --- {\it modular polarization} (defined via the Zak transform). Given Schr\"{o}dinger's $\psi_n(x)$ define the modular wave function
\begin{equation}
\psi_{\lambda} (x, \tilde{x}) \equiv \sqrt{\lambda} \sum_n e^{-2\pi i n \tilde{x}} \psi_n (\lambda(n+x)),
\end{equation}
($x \equiv q/{\lambda}$, $\tilde{x} \equiv p/{\epsilon}$, so $[x, \tilde{x}]=i$, $\lambda \epsilon = \hbar$).
From the point of view of modular polarization, Schr\"{o}dinger's polarization is very singular.
Introduce $\X^{A}\equiv (x^a , \tilde{x}_a)^{T}$, so that $ [ \hat{\X}^a, \hat{\X}^b] = i \omega^{AB}$.
We can write the translations operators in phase space covariantly $W_{K} \equiv e^{2 \pi i \omega( K, X)}$,
where $K$ stands for the pair $(\tilde{k},k)$ and  $\omega(K,K')=k \cdot \tilde{k}' - \tilde{k}\cdot k'$.
($W$ should be really understood as {Aharonov-Bohm phases, which are prototypical examples of modular variables}.)

So far we have discussed covariant quantum phase space as an examples of modular space, and so we are ready to discuss modular spacetime.
Consider\cite{Freidel:2018apz} 
a {\it metaparticle} (mp) propagating in a modular space defined by Born geometry, $(\omega, \eta, H)$. The metaparticle world-line action
$S_{mp} = \int d\tau L_{mp}$ (with the canonical particle emerging in the $\mu \to 0$ and $\tilde{p} \to 0$ limit), with
$L_{mp} = p_\mu\, \dot x^\mu +\tilde p^\mu\, \dot{\tilde x}_\mu +\lambda^2 p_\mu\, \dot{\tilde p}^\mu - 
\frac{N}{2}\left(p_\mu p^\mu +\tilde p_\mu \tilde p^\mu - m^2\right) +{\tilde{N}}\left(p_\mu \tilde p^\mu - \mu \right),$
where $\omega$ is in (``the Berry-phase'') $p_\mu\, \dot{\tilde p}^\mu$, and $\eta$ in the diffeomorphism constraint $p_\mu \tilde p^\mu = \mu$
and $H$ in the Hamiltonian constraint $p_\mu p^\mu +\tilde p_\mu \tilde p^\mu = m^2$.
Here it is natural to talk of {\it dual spacetime} $\tilde{x}$, $[x, \tilde{x}] = i \lambda^2$, and {\it dual momentum space} $\tilde{p}$, $[p, \tilde{p}] = 0$.
(Also, $[x, p] = i\hbar = [\tilde{x}, \tilde{p}]$.)

The metaparticle can be understood also as follows: If one second quantizes Schr\"{o}dinger's $\psi(x)$ one naturally ends up with a quantum field operator ${\hat{\phi}}(x)$.
Similarly, the second quantization of the modular $\psi_{\lambda} (x, \tilde{x})$ would lead to
a modular quantum field operator ${\hat{\phi}}_{\lambda} (x, \tilde{x})$ ({\it modular fields} or {\it metafields})
\be
{\hat{\phi}}(x) \to {\hat{\phi}}_{\lambda} (x, \tilde{x}),
\ee
with $[x, \tilde{x}] = i \lambda^2$ defining a covariant non-commutative field theory \cite{Freidel:2017wst, Freidel:2017nhg} that depends on the contextuality parameter $\lambda$.
Classical spacetime label $x$ of canonical QFT corresponds to a choice of (classical spacetime) polarization in modular (quantum) spacetime with 
a contextuality parameter $\lambda$.
Quanta of canonical quantum fields $\phi(x)$ are particles (and their antiparticles). Similarly,
{quanta of modular quantum fields $\phi_{\lambda}(x, \tilde{x})$ are metaparticles.}
{Thus, the first prediction of modular spacetime approach to quantum theory concerns the existence of metaparticles}\cite{Freidel:2018apz}. 
(We will argue that dual particles, correlated to visible particles, represent dark matter.)
If we turn on backgrounds $p \to p + \phi$ and $\tilde{p} \to \tilde{p} + \tilde{\phi}$.
Thus we have ``dark matter'' fields, $\tilde{\phi}(x)$, in the effective classical spacetime $x$ description (after integrating over
the dual spacetime $\tilde{x}$).
{However, the visible $\phi$
and invisible (dark matter) ${\tilde{\phi}}$ do not commute.}
Therefore, one could say that dark matter is fuzzy from the point of view of classical spacetime.

\subsection{Some Consequences of Modular Spacetime}
\label{s:ModFlux}
A few other comments are in order\cite{Freidel:2016pls, Freidel:2015pka, Freidel:2017xsi}:
{\it modular spacetime has double the dimension of spacetime}.
The modular cells are not simply connected (there is a unit flux through each cell --- this could be considered as the origin of matter/fermionic degrees of freedom; the deformations of each cell could be understood as the origin of bosonic degrees of freedom, i.e. interaction quanta). We note the possibility of general quantum statistics and general higher order quantum-correlations, to be discussed in Section~\ref{s:EspGQ}. Similarly, one needs to adopt double scale (UV/IR) Renormalization Group characteristic of non-commutative field theory \cite{Freidel:2017xsi}
in order to define the continuum limit as well as renormalization of couplings and correlation functions in the context of modular quantum field theory.
In some sense, modular quantum field (metafield) theory is the limit of the underlying quantum gravitational origin of quantum field theory defined in quantum spacetime, with the local (Wilsonian) quantum field theory in a fixed spacetime background, being a singular, classical spacetime limit, of this more fundamental description.

{We emphasize that modular wavefunctions are quasiperiodic} \cite{Freidel:2016pls}.
Classical spacetime emerges from the process of {\it extensification} (imagine one unit length in dual direction
and many, $N$, modular cells in the spacetime direction).
Spacetime emerges, in the large $N$ limit, as a natural pointer basis in quantum theory \cite{Freidel:2016pls}.
(And this sheds new light on the problem of quantum measurement. The dual spacetime labels can be understood as covariant ``hidden variables'' associated with the spacetime pointer basis, selected by quantum gravity.)
Also, {\it spacetime and matter appear as `two sides of the same coin''}.
{Similarly, cosmology is an interplay between visible and dual (invisible) degrees of freedom \cite{Freidel:2021wpl}.}

One can compute the propagator for the metaparticle\cite{Freidel:2018apz} 
\begin{equation}\label{doubletramp}
G(p,\tilde p; p_i,\tilde p_i)
\sim
\delta^{(d)}(p-p_{i})\delta^{(d)}(\tilde p-\tilde p_{i})
\frac{\delta(p\cdot\tilde p-\mu)}{p^2+\tilde p^2+m^2-i\varepsilon}.
\end{equation}
The canonical particle propagator is a highly singular $\tilde p \to 0$ (and $\mu \to 0$) limit of this expression.
One also obtains the following dispersion relation (in a particular gauge $\vec{\tilde{p}} = 0$)
\begin{equation}
  E_p^2 + \frac{\mu^2}{E_p^2} = \skew{-2}\vec{p}^2 + m^2.  
\end{equation}
For each particle at energy $E$ there exists a dual particle at energy $\frac{\mu}{E}$.
(Analogous to the prediction of antiparticles in QFT.)
This dispersion relation is indicative of quantum gravity phenomenology in the infrared limit, thus contradicting the usual intuition about the relevance of the Planck scale for quantum gravity. 
(Given the relevance of quantum field theory in the context of many body physics, one might wonder if metafields and metaparticles find their use in that domain of physics as well. Indeed, one can consider non-relativistic metafields and their metaparticle quanta and one can, for example, derive a Friedel-like static potential for metaparticles and also
introduce the concept of quasi-metaparticles in the realm of many body physics \cite{Barnes:2021akh}.)

{Note that dual ``particles'' (dual fields) are natural candidates for dark matter}  because to leading order in $\lambda$
\begin{equation}
    \Seff = - \int \sqrt{{g(x)}{\tilde{g}(\tilde{x})}} [R(x) + \tilde{R}(\tilde{x})+ 
    L_m (A(x, \tilde{x})) + \tilde{L}_{dm}   ( \tilde{A}(x, \tilde{x})) ].
\end{equation}
Here the $A$ fields denote the usual Standard Model fields, and the $\tilde{A}$ are their duals, as predicted by the general (modular) formulation of 
quantum theory that is sensitive to the minimal length.
In the above expression one needs to integrate over the dual space coordinates $\tilde{x}$ to get an effective
description of {\it visible matter, $A(x)$,  and dark matter, $\tilde{A}(x)$}, in classical $x$ spacetime.

{Similarly, dynamical geometry of dual spacetime represents the natural origin of dark energy} (again, to leading order in $\lambda$) 
\begin{equation}
    \Seff = - \int \sqrt{-g(x)} \sqrt{-\tilde{g}(\tilde{x})} [R(x) + \tilde{R}(\tilde{x})+\dots].
\end{equation}
In this leading limit, the $\tilde{x}$-integration in the first term 
defines the gravitational constant $G_N$, and in the second term produces a {positive cosmological constant constant}!
{In general, visible and dark matter degrees of freedom are correlated (via the minimal length $\lambda$). This suggest the origin (from dark matter) of the observed scaling (found in galaxies, clusters, superclusters) and the universal acceleration} $a_0 \sim c H/(2 \pi)$
(with the observed positive cosmological constant $\Lambda \sim H^2$).
Metaparticles can be understood as fuzzy dark matter, which in turn does point to a natural relation of vacuum energy (dark energy) and fuzzy dark matter \cite{Edmonds:2017zhg, Edmonds:2024qsj}.

\subsection{Dynamical Born Geometry: Realization in String Theory} 
\label{s:mStr}
As already emphasized in Section~\ref{s:CC-fix}, the above phase space/modular formulation is naturally realized in terms of a chiral, phase-space-like, intrinsically non-commutative and T-duality covariant reformulation of the bosonic string, the {\em metastring}\cite{Freidel:2013zga,Freidel:2015uug,Freidel:2015pka} (which may also be turned into a non-perturbative proposal\cite{rBHM10,Berglund:2021hbo,Berglund:2022qsb}
a matrix model-like, time-asymmetric (in general), $\partial_{\sigma} \cdot \equiv [\hat{\X}, \cdot]$, where $\hat{\X}$ matrix comes from the modular world-sheet); here the matrix entries correspond to spacetime/matter quanta (``monads'')):
\begin{equation}
S^{\text{ch}}_{\text{str}}=
\int\rd \tau\,\rd \sigma~
    \Big[\pa_{\tau}{\X}^{a} \big(\eta_{ab}(\X)+\omega_{ab}(\X)\big)
    -\partial_\sigma\X^a H_{\!ab}(\X)\Big] \partial_\sigma\X^b. 
\end{equation}
Here, $\X^a\define (X^a/\ls ,\tilde X_a/\ls )^{T}$ are
coordinates on phase-space like (doubled) target spacetime and the fields $\eta, H,\omega$ are all dynamical (i.e., generally $\X$-dependent) target spacetime fields\footnote{Recall that the ten-dimensional superstring comes out of the twenty-six dimensional bosonic string\cite{Casher:1985ra}. Similarly, the matrix formulation of M-theory in eleven dimensions would emerge from a non-perturbative matrix formulation of the metastring\cite{Berglund:2022qsb} (notice the lack of the overall trace --- this is a matrix model as matrix quantum mechanics of Born-Heisenberg-Jordan), and it should be understood as gravitized quantum theory,
$
{\S}^{\text{non}}_{\text{str}}=
\int\rd \tau\;
     \Big(\big[\pa_{\tau}{\hat{\X}}^{A}  g_{ACD}(\hat{\X}) 
    - [\hat{\X}^A, \hat{\X}^B]h_{ABCD}(\hat{\X})\big] 
    [\hat{\X}^C, \hat{\X}^D] \Big),
$
where $A,B,C,D$ run from $0,1,\dots 26$. Note that the first term is Chern-Simons-like and the second Yang-Mills-like, both being of the open string origin. The fully compactified non-perturbative bosonic metastring provides the basic elementary constituents for {\em extensifications}\cite{Freidel:2015pka, Freidel:2016pls}  from zero dimensions to four quantum spacetime dimensions (natural from the point of view of string cosmology\cite{Brandenberger:1988aj}).
Finally, we note an interesting structural connection of the non-perturbative metastring to the ``palatial'' (non-commutative) twistor theory \cite{Penrose:2015lla}. The intrinsic time asymmetry of the non-perturbative metastring formulation might be the origin of an intrinsic gravitational $CP$ violation.}.
In terms of the left- and right-moving 0-modes of the twenty-six dimensional bosonic string, one defines
\begin{equation}
    x^a\define x^a_L + x^a_R,\quad \Tilde{x}^a \define \Tilde{x}^a_L - \Tilde{x}^a_R.
\end{equation}
In the context of a {\em flat\/} metastring, the coefficients
 $\eta_{ab}$,  $H_{ab}$
 and $\omega_{ab}$\footnote{Setting $\omega_{ab}=0$ directly connects $S^{\text{ch}}_{\text{str}}$ to double field theory.} are constant:
\be\label{etaH0} 
	\eta_{ab} = \left( \begin{array}{cc} 0 & \delta \\ \delta^{T}& 0  \end{array} \right),\quad
H_{ab} =  \left( \begin{array}{cc} h & 0 \\ 0 &  h^{-1}  \end{array} \right),
\quad \omega_{ab} = \left( \begin{array}{cc} 0 & \delta \\ -\delta^{T}& 0  \end{array} \right) ,
\ee
where $h$ denotes the flat $(1,d{-}1)$-dimensional metric and $\d$ is the Kronecker symbol.
The standard Polyakov action is then obtained by setting $\omega_{ab}=0$ and integrating out the $\tilde x_a$,
\begin{equation}
    S_P=\int \rd \tau\,\rd \sigma~ \gamma^{\alpha\beta}\, \partial_\alpha X^a\, \partial_\beta X^b\, h_{ab} + \ldots
\end{equation}
The triplet $(\omega,\eta,H)$ defines Born geometry\cite{Freidel:2013zga,Freidel:2015pka} (which is ultimately dynamic, suggesting a ``gravitization of quantum theory''\cite{Freidel:2014qna,Berglund:2022qcc,Berglund:2022skk}) so that the metastring  propagates in a (dynamical) modular spacetime, a phase space like structure that naturally arises in any quantum theory\cite{Freidel:2016pls}, as argued in the previous subsection. One of the key consequences of this is that the metastring is intrinsically non-commutative and also that its low energy QFT-like description in modular spacetime is intrinsically non-commutative.
Thus every Standard Model field $\phi(x)$ is doubled as $\phi(x, \tilde{x})$ and $\tilde{\phi}(x, \tilde{x})$,
with doubled and non-commutative arguments
$[x^a, {\tilde{x}}_b] = i\ls^2\, \delta^{a}{}_{b}$.
The quanta of such modular fields are the zero modes of the metastring --- the {\em metaparticles} --- whose dynamics, as already discussed, are given by a world-line action involving a doubling of the usual phase space coordinates. The metaparticle (``mp'') action describes the zero modes of the metastring and is of the form
which is fixed by the three ingredients of the Born geometry
\cite{Freidel:2017wst, Freidel:2017nhg, 
Freidel:2018apz, Freidel:2021wpl}
\begin{equation}\label{mp1}
S_{\text{mp}} \define \int_0^1 d\tau \Big[p\cdot \dot x +\tilde p\cdot \dot{\tilde x}+ \alpha'\, p \cdot\dot{\tilde p}
- \frac{N}2\left(p^2 +{\tilde p}^2 + \mathfrak{m}^2\right) +{\tilde N}\left(p\cdot \tilde p - \mu \right)\Big]\,,
\end{equation}
where the dot-product denotes contraction with signature $(-,+,\ldots,+)$.
The new feature here is the presence of a non-trivial symplectic form on the metaparticle phase space, the non-zero Poisson brackets being
\beq
\{p_\mu, x^\nu\}=\delta_\mu^\nu,\quad
\{\tilde p_\mu, \tilde x^\nu\}=\delta_\mu^\nu,\quad
\{ \tilde x_\mu,x^\nu \}= \pi\alpha'\,\delta_\mu^\nu,\quad
\alpha'\sim\ls^2.
\label{xtxcomm}
\eeqn
with $\mu,\nu=0,1,\dots,d-1$\footnote{The dual momentum $\tilde p$
could be understood from the point of view on conserved (electric) charges and corner symmetries in quantum gravity\cite{Strominger:2017zoo, Freidel:2020xyx, Freidel:2020svx, Ciambelli:2021vnn}. The dual spacetime 
$\tilde x$ is associated with conserved dual (magnetic) charges in the same context.}.
Because of its interpretation as a particle model on Born geometry, associated with the modular representation of quantum theory,
the space-time on which the metaparticle propagates is ambiguous, with different choices related by what in string theory we would call T-duality.
The attractive features of this model include world-line causality and unitarity, as well as an explicit mixing of widely separated energy-momentum scales. (Note that the Kalb-Ramond 2-form can be naturally incorporated in the non-commutativity of dual spacetime coordinates. Thus the non-commutativity (and in general, non-associativity consistent with the projective geometry of quantum theory) of the closed string is already present in the massless spectrum of the string. Similarly the Kalb-Ramond field, and its T-dual, allow for manifest relative locality in string theory, as they imply the generic mixing between the spacetime coordinates and their dual counterparts\cite{Freidel:2017wst, Freidel:2017nhg}. Therefore, the classical spacetime of EFT is just one polarization allowed in the more fundamental description, and it is tied to the contextual nature of quantum measurement.)

The non-commutative algebraic structure of the effective metaparticle description (fixed by Born geometry) is realized in metastring theory\cite{Freidel:2013zga,Freidel:2015uug,Freidel:2015pka}, merely with ``softening'' the indeterminacy by replacing, $l_s\to l_{eff}$, where 
$l_{eff}$ is the relevant effective length-scale in~\eqref{xtxcomm}.
The central point here is that the metastring formulation explicitly realizes modular spacetime and the modular polarization needed in the argument for the bound of the cosmological constant, and, as a theory of quantum gravity, also realizes the Bekenstein bound.
Thus, a natural realization of the resolution of the cosmological constant problem is found in the metastring formulation of string theory.
Moreover, string theory is a quantum theory of gravity {\em and\/} Standard Model-like matter. Therefore, other vexing problems beside the cosmological constant problem should be possible to address in the same context, to wit, the gauge hierarchy problem and the problem of fermion masses and mixing angles.
This will be done in Section~\ref{s:PPP}.

{How could we ``observe'' modular spacetime?}
Instead of scattering particles, we could entwine them.
The canonical vertex operators $\widetilde{V}_{P}$ (plane waves  or asymptotic particle states) have co-cycles in the Polyakov string if we
assume that $[x, \tilde{x}] =0$ \cite{Polchinski:1998rq}
\be
\widetilde{V}_{P} \widetilde{V}_{P'}
= e^{i (p\tilde{p}' -\tilde{p}p')}\widetilde{V}_{P'} \widetilde{V}_{P}.
\ee
The cocycle factor $e^{i (p \tilde{p}' - \tilde{p} p')}$ indicates the fundamental non-commutativity
of $x$ and $\tilde{x}$.
Can this entwining of particles be measured?
(Here we are really talking about the ``R-matrix'', in the sense of ``swapping of particles'', 
instead of
the S-matrix i.e. ``scattering of particles''.)
Note, that the free energy of the metastring scales, as it
should by the old argument of\cite{Atick:1988si}, as a 2d field theory (the world sheet CFT). This old observation is consistent with the above non-perturbative formulation of metastring theory. Thus one could search for this
fundamental dimensional reduction, consistent with ``asymptotic silence'' in general relativity\cite{Belinskii:1971Osc}. 
At some critical energy the observed jets should be planar (here one expects a phase transition between the generic non-planarity of the Standard Model jets below some critical energy and the generic planarity of jets above that critical energy\cite{Anchordoqui:2010er}\footnote{
Work in progress by Nikolina Ilic, Dejan Stojkovic, Doug Gingrich, Luca Colangeli, Mathias Roman, Sebastien Roy-Garand, Yun Qing Wu and DM.}.

\section{Experimental Probes of Gravitized Quantum Theory}
\label{s:EspGQ}
In this section we discuss the experimental signature of quantum gravity viewed as gravitized quantum theory.
In particular, we discuss higher order quantum interference effects\footnote{We thank Per Berglund, Andy Geraci and Dave Mattingly for conversations on this topic.} and we also comment on the statistics of spacetime quanta and the relevant
experimental probes of the same.

\subsection{Gravitized Quantum Theory: Top-Down and Bottom-Up}

The metastring has {dynamical Born geometry}, $\omega_{ab}(\X),\eta_{ab}(\X), H_{ab}(\X)$, 
but Born geometry is the geometry of the modular spacetime formulation
of quantum theory. 
Thus {by making Born geometry dynamical we can ``gravitize quantum theory''} (that is, {make the geometry
of quantum theory dynamical}) \cite{Freidel:2014qna, Freidel:2013zga}.
The metastring is a theory of quantum gravity, and so we arrive at the advertised dictionary {``quantum gravity = gravitized quantum theory''.}{ In what follows, we argue that triple and higher order quantum interference is one of the central observational consequences of this dictionary.}
This reasoning is ``top-down''.

Recall that classical classical gravity (general relativity) gravitizes all of classical physics, but making the relevant fundamental equations generally covariant, and thus it
is not strange to expect that quantum gravity requires gravitization of quantum theory.
Therefore, consider particle interactions as $0{+}1$ quantum gravity, by 
remembering that quantum field theory can be understood as $0{+}1$ quantum gravity/cosmology. (This discussion could be generalized for the case of string field theory and its fundamental cubic vertex, the ``pants diagram'', by viewing string theory, including its metastring formulation, as 
$1{+}1$ quantum gravity/cosmology.) 
For example: consider the $\phi^3$ theory. The relevant classical equations read
\be
(\partial^2 + m^2) \phi + g \phi^2 =0
\ee
{Here we should understand the classical field $\phi$ as the wave-function of $0{+}1$ universes.}

Thus the above classical field equation should be understood as a non-linear Wheeler-DeWitt (WdW) equation.
The interaction vertex is indicative of {topology change}.
We have the classical spacetime viewpoint where the decay of a particle (or scattering of particles, the S-matrix)  is
viewed by utilizing the Born rule.
However, the QG=GQ viewpoint from $0{+}1$ universes (particles) is that there
exists an intrinsic triple correlation. This is another motivation for ``gravitization'' of quantum theory. For example, the fundamental triple vertex (the pants diagram) of string theory becomes the fundamental triple quantum correlation of $1+1$ dimensional quantum gravity. (Also, such gravitization of quantum theory, in the context of the metastring formulation of string theory, sheds new light on the background independent definition of string field theory.)

In general, if we view general relativity as a theory of interacting gravitons
(closed strings), we have {$n$-correlations; $n = 3,4,5\dots$} and thus the fundamental object of general relativity, the classical spacetime, hides higher order (and in principle infinite order) quantum correlations. By reinterpreting the non-linear interactions terms in the non-linear WdW equation as the time evolution operator, time could be understood as measuring the rank of general quantum correlations, and space as the size of those correlations. This sheds a new light on the problem of time in quantum gravity, as well as at the fundamental question of the expanding spacetime in the context of cosmology. The fundamental quantum gravity description contains all quantum correlations, and the emergent classical spacetime hides all correlations except for the maximally symmetric (Born-like) correlations associated with the matter degrees of freedom. Thus quantum matter that exists in classical spacetime follows the Born rule and canonical quantum theory. The classical world, or the results of quantum measurement, represents the memory of those higher order quantum correlations (topology change in the space of quantum states) hidden in the classical spacetime ``condensate'' in which the canonical quantum theory operates, with definite results of quantum 
measurements in such classical spacetime.

Apart from the top-down rationale, there exists a ``bottom-up" reason for ``gravitization of quantum theory'' \cite{Minic:2003en, Minic:2003nx,
 Minic:2004rj, Minic:2002pd}.  
{The canonical geometry of quantum theory} (as reviewed by Ashtekar and Schilling
\cite{Ashtekar:1997ud} as well as\cite{Jejjala:2007rn}) is encoded in a maximally symmetric geometry of complex projective spaces
(defined by a symplectic structure, compatible with the metric structure --- the Born rule --- and the product of the
symplectic and metric structure that defines the complex structure, ultimately responsible for quantum interference).
The quantum clock relates the Born rule (the Fubini-Study metric of complex projective spaces) to infinitesimal time\cite{Anandan:1990Geo}
(see also Aharonov's earlier work\cite{Aharonov:1987Pha}),
\begin{equation}
    2 \hbar \,\rd s_{FS} = \Delta E\, \rd t,
\end{equation}
where $\Delta E$ is the dispersion of energy defined by the Hamiltonian associated with the atomic clock.

In the presence of quantum spacetimes (where topology change is allowed) there exists no unique timelike Killing vector and thus $\Delta E$ is state dependent
which makes the geometry state dependent, and thus, dynamical. 
{One can also recall that a dynamical inner product occurs in the context of 2+1 quantum gravity viewed as a Chern--Simons theory\cite{Witten:1989ip}. 
So, for quantum spacetimes, we should expect ``gravitized quantum theory'', 
that is, {a dynamical geometry of quantum theory},
and, in general, we should allow for topology change in the space of states.
(Thus the Bloch sphere becomes a Riemann surface of an arbitrary genus.)
Hence, even from the bottom up approach, both the geometry and topology of the space of quantum states should become dynamical in quantum gravity viewed as a gravitized quantum theory.

In general, we should not expect to have a single Hilbert space in quantum gravity, but {observer dependent Hilbert spaces, implied by the already emphasized existence of observer dependent spacetimes}, and thus we should expect a generalization of the usual kinematics of quantum theory (which is consistent with a matrix-like formulation of non-perturbative metastring theory in section 3).
In what follows we will argue that a dynamical Born rule and such generalized kinematics imply new experimental signatures (triple and higher order
quantum interference). This requires the introduction of new observables that go beyond the S-matrix, decay rates, transition amplitudes, etc.
(In other words, quantum theory with a single Hilbert space requires the Born rule with canonical observables, like the S-matrix. In general, we should expect observer dependent Hilbert spaces and higher order quantum interference and more general observables.)

\subsection{Higher Order Interference and Quantum Gravity}

What is the first experimental consequence of ``gravitized quantum theory''? 
{\it We claim the following\cite{Berglund:2023vrm}: The experimental ``smoking gun'' of gravitized quantum theory is represented by triple and higher-order interference of matter waves in a gravitational background.} (This experiment is in principle possible in the next few years.)

The canonical quantum theory does not have intrinsic triple quantum interference (as a consequence of the Born rule
and the fixed geometry of the complex projective space).
{ Current experimental (photonic) bounds on triple interference are rather weak ($10^{-3}$). Neutrino bounds are expected to be
surprisingly similar (and to be measured at JUNO
\cite{Huber:2021xpx}).}
In more detail\cite{Sorkin:1994dt}: Classically, we have
addition of probabilities
\be
P_{n}(A,B,C,\cdots) \,=\, P_{1}(A) + P_{1}(B) + P_{1}(C) + \cdots\;,
\ee
for any number of paths.
Quantum mechanically, we have for two paths
$P_{2}(A,B) = |\psi_A + \psi_B|^2 \vphantom{\Big|}$,
or more explicitly
\be
{|\psi_A|^2} + 
{|\psi_B|^2} + 
{(\psi_A^*\psi_B^{\phantom{*}} + \psi_B^*\psi_A^{\phantom{*}})}
\equiv P_{1}(A) + P_{1}(B) + I_{2}(A,B),
\ee
where the last term
\be
I_{2}(A,B) = P_{2}(A,B)-P_{1}(A)-P_{1}(B),
\ee
is the ``interference'' of the two paths $A$ and $B$.
{\it Thus, the non-vanishing double-path interference, 
$I_{2}(A,B)\neq 0$, distinguishes quantum theory from the classical one.}

The {Born rule} dictates that all the superimposed paths only interfere with each other 
in a pairwise manner.
For instance, for three paths we have
$P_{3}(A,B,C) = |\psi_A {+} \psi_B {+} \psi_C|^2$
\be
P_{2}(A,B) {+} P_{2}(B,C) {+} P_{2}(C,A) 
{-} P_{1}(A) {-} P_{1}(B) {-} P_{1}(C),
\label{eq:three-slit}
\ee
where only pairwise interferences between the pairs $(A,B)$, $(B,C)$, and $(C,A)$ appear.
It is clear from the above that in order for there to be a non-linear correction in an interference pattern the Born rule must be relaxed.  
Consider a triple slit experiment:
Since only pairwise interferences between the pairs $(A,B)$, $(B,C)$, and $(C,A)$ appear, it makes sense to define any deviation from this relation as the intrinsic
triple-path interference $I_{3}(A,B,C)$ 
\be
P_{3}(A,B,C)
-P_{2}(A,B)
-P_{2}(B,C)
-P_{2}(C,A)
+P_{1}(A)
+P_{1}(B)
+P_{1}(C).
\label{e:I3}
\ee
(This can be easily generalized for the case of $n$-paths.)
For both classical and quantum theory, this intrinsic triple-path interference is zero for any triplet of
paths. {Experimental confirmation of $I_3=0$ would be a confirmation of the Born rule.}  
Weak bounds were placed on the parameter ($\kappa \sim 10^{-3}$)
in photonic experiments (see the references in \cite{Berglund:2023vrm})
\be
\kappa = \dfrac{\varepsilon}{\delta}, \quad \varepsilon =  I_3(A,B,C), \quad \delta  =  |I_2(A,B)| + |I_2(B,C)| + |I_2(C,A)|.
\ee

The claim of\cite{Berglund:2023vrm}  
is that {\it with quantum gravitational degrees of
freedom turned on, one can get $I_3 \neq 0$,}
but for that one needs gravitized quantum theory, with
{\it observer dependent Hilbert spaces and dynamical Born rule}.
Inspired by metastring theory,
the generalized probability in this approach to quantum gravity is given by
\be
  P =    g_{ab}(\psi)\, \psi_a \psi_b \equiv \delta_{ab}\, \psi_a \psi_b + \gamma_{abc}\, \psi_a \psi_b \psi_c+\dots
   ,
\label{e:deformP}
\ee
where $a,b,c$ are state-space indices and with (schematically).
One non-relativistic quantum gravity model is provided by the canonical Schr\"{o}dinger dynamics perturbed by Nambu quantum theory\cite{Berglund:2023vrm} (in the non-relativistic limit)
\be
\frac{\rd \psi_a}{\rd \tau} = \Gamma_{abc}\, \psi_b \psi_c,
\label{e:deformDpsi}
\ee
where $\tau$ is the appropriate evolution parameter (and higher order generalizations $\frac{\rd \psi_a}{\rd \tau} = \Gamma_{abcd}\, \psi_b \psi_c \psi_d$, etc.
Here $\Gamma_{abc}$ is such that one has Schr\"{o}dinger's evolution
for a fixed Hilbert space.) Note that the Schr\"{o}dinger equation can be understood as a geodesic equation on complex projective spaces, which are
Einstein's spaces (maximally symmetric, homogeneous and isotropic). The above generalized evolution can be understood to originate from the self-dual nature of the equations of motion of the metastring\cite{Freidel:2013zga} (a fully relativistic gravitized quantum theory). Thus even the generalized quantum evolution equation can be understood as a geodesic equation of more general Einstein-like equations on the spaces of quantum states with general geometry and topology. These in turn are consistent backgrounds of the non-perturbative metastring endowed with fully dynamical Born geometry.

Here we comment that the above model of Nambu quantum theory\cite{Minic:2002pd, Minic:2020zjb} is essentially
based on volume preserving diffeomorphisms where the generator of volume preserving transformations is caused by the
Nambu bracket\cite{Nambu:1973qe}, a generalization of the Poisson bracket.
The classic example is the asymmetric top which can be re-written as a model of Nambu's classical mechanics.
The Nambu quantum theory can be understood in the above Schr\"{o}dinger representation or in the
matrix representation which requires cubic and higher order matrices\cite{Awata:1999dz}.
In gravitized quantum theory complex projective spaces are
generalized (for example: the Bloch sphere becomes a Riemann surface of infinite genus) where higher order quantum correlations are indicative of a dynamical Born rule (with handles representing higher components $\psi_3$, where $\psi_1$ and $\psi_2$ are the real and imaginary parts of the canonical complex wave function).
The classical limit is captured by topological branching (the quantum metric, or, equivalently, probability, becomes degenerate and equal to zero).
A re-summation of the infinite number of multilinear extensions results in a general-relativity-like theory
in the general space of states (that includes $\psi_3$ etc).
The canonical quantum theory is the maximally symmetric limit of this more general and ``gravitized'' formulation of quantum theory (emerging by averaging over the infinite number of handles).

{Notice that effective triple interference is possible in non-linear optical media\cite{Namdar:2021czo} and this experiment provides another motivation for our work\cite{Berglund:2023vrm}.}
(In that context, instead of $\psi$  we have non-linear waves and instead of probability $P$ --- non-linear/cubic energy density.)
The ``smoking gun'' experiment for gravitized quantum theory is thus the
{\it Talbot effect on a diffraction grating that is turned into a non-linear Talbot effect}\cite{Berglund:2023vrm}.
This intrinsic triple interference with quantum gravity degrees of freedom is analogous to the model of quantum spacetime as a non-linear ``quantum spacetime medium''.
(Here we stress that {\em no} non-linear quantum theory with {\em fixed Hilbert spaces} can possibly contend to be a gravitized quantum theory.)
Based on the discussion of the vacuum energy/cosmological constant (possibly the first experimentally detected effect of quantum gravity)
it can be argued that quantum gravity effects appear at low energy scales: 
such as the cosmological constant scale {$10^{-4}$\,m or or the natural particle physics scale $10^{-19}$\,m.}\footnote{Born rule is also used as the entanglement witness in the experimental probes of the quantum nature of gravity, which involve entanglement of masses --- Quantum Gravity via the Entanglement of Masses (QGEM) \cite{Bose:2017nin, Marletto:2017kzi,
Belenchia:2018szb}. According to our general reasoning, even in this context one should check the validity of the Born rule, and expect possible deviations from it.}

\subsection{Infinite Statistics and Spacetime Quanta}
In this subsection we want to address the question of statistics of spacetime atoms as implied by our solution of the vacuum energy problem, to which end we start from the following observation about the atomic structure of matter.
Matter is granular and cuttable: it consists of fermions that
are held together through interactions that are mediated by bosons (the spin-statistics theorem of local QFT). In order to extend this atomic picture to spacetime we note the fundamental difference between spacetime and matter:
spacetime is extended and non-cuttable. In what follows we claim that spacetime quanta obey infinite statistics and are held together by higher order quantum correlations responsible for higher order interference effects (already discussed in the previous subsection). 
Thus, classical spacetime (with all its features) is a left-over from the quantum gravity phase, which involves all higher order (triple and higher) quantum correlations.
On the other hand, matter is captured by degrees of freedom with only Born-like
quadratic and maximally symmetric correlations of canonical quantum theory in a background
of such an emergent classical spacetime. Classical spacetime also provides the pointer basis for quantum measurements involving matter degrees of freedom, with an inevitably classical nature of such measurement outcomes.

We argue\footnote{We thank Laurent Freidel, Jerzy Kowalski-Glikman and Rob Leigh for conversations about this topic.} that the quanta of quantum spacetime
defined as modular spacetime obey infinite statistics.
Using infinite (or quantum Boltzmann) statistics we derive the fluctuation of (modular) energy
discussed by Verlinde and Zurek  
\cite{Verlinde:2019ade}
and by Zurek in
\cite{Zurek:2020ukz}.
In what follows we will use the basic formulae of
the thermodynamics of infinite statistics from Section~\ref{s:QG=GQ} of Ref.  
\cite{Minic:1997ym}, and references therein (see also, \cite{Ho:2012ar, Jejjala:2020lhg}).

Given the fundamental commutator, $[x, \tilde{x}]=i l_P^2$, 
between spacetime $x$ and dual spacetime $\tilde{x}$, discussed in Section~\ref{s:QG=GQ},
both $x$ and $\tilde{x}$ have
to be infinite matrices. Thus their probability distribution has
be governed by non-commutative probability theory or
equivalently by quantum distinguishable, or quantum Boltzmann
statistics, also known as infinite statistics.
This statistics (which, in the simplest case, is
the analog of the Gaussian statistics for non-commutative variables)
and which is studied as such in non-commutative probability
theory of Voiculescu and collaborators \cite{free} and reviewed in the paper
by Gross and Gopakumar \cite{Gopakumar:1994iq} can be applied to modular cells of modular spacetime, which are quantum and distinguishable.

Note that Strominger \cite{Strominger:1993si} pointed out that the concept of infinite  statistics is relevant for black holes,
and this fits naturally with the claim that the statistics of “spacetime atoms”, as 
represented by modular cells, is quantum statistics of distinguishable objects.
Modular cells are quantum and they are distinguishable by construction, and thus it is not surprising that they should obey quantum distinguishable or quantum Boltzmann statistics, that is, infinite statistics, the natural covariant statistics of extended objects. In fact, Greenberg \cite{Greenberg:1989ty, Greenberg:1991ec} has pointed out that infinite statistics is the only statistics consistent with non-locality and CPT/Lorentz symmetry. This again is consistent with the basic features of modular spacetime cells ---
modular spacetime is covariant, but modular cells are non-local objects.
(Also, modular spacetime is the background of the metastring formulation of string theory, a covariant theory of non-local objects --- strings. In principle, it should be possible to prove a stringy version of spin-statistics theorem in string field theory, which would naturally lead to infinite statistics of strings.)

In what follows we present a summary of basic concepts from
\cite{Minic:1997ym}
(and the references therein).
Infinite statistics is defined in terms of
a free algebra (Cuntz algebra) of operators
\be
a_i a_j^{\dagger} =\delta_{ij},
\label{e:Cuntz}
\ee
where, as usual, $a$ annihilates the vacuum
\be
a_i |0\rangle =0.
\ee
Note that, unlike the Bose-Einstein (BE) or Fermi-Dirac (FD) statistics for matter (bosonic or fermionic) degrees of freedom, we do not have a commutator or anticommutator relation between $a$ and $a^{\dagger}$.
The defining equation~\eqref{e:Cuntz} is the exceptional $q=0$ case of the $q$-deformed (quon) commutator 
\be
a_i a_j^{\dagger} - q a_j^{\dagger} a_i = \delta_{ij},
\ee
where BE corresponds to $q=1$
and FD to $q=-1$, in both of which quanta are indistinguishable.
In contradistinction, the exceptional $q=0$ case specifies the algebra~\eqref{e:Cuntz} for a quantum Boltzmann statistics, that is, a {\em distinguishable}
quantum statistics. 

The thermodynamics of a system of particles
obeying distinguishable quantum statistics was studied in \cite{Liu:1997gk} (sections 3 and 4).
These calculations were motivated by the
discussion of Schwarzschild black holes as bound
states of $D0$-branes in Matrix theory, a light-cone, holographic, formulation of M-theory
\cite{Banks:1996vh}, 
as explored in \cite{Banks:1997hz}.
The partition function $Z =\sum_i e^{-\beta E_i}$ of a free gas (in $d-2$
transverse dimensions of spacetime) of
$N$ particles obeying infinite statistics 
with the leading order free Lagrangian $L = \frac{N}{R} v^2$ is
\cite{Liu:1997gk} 
\be
Z = (V)^N (T/R)^{N (d-2)/2},
\ee
where $V$ is the volume and $T\sim \frac{1}{\beta}$ temperature, and $R$ is a characteristic length scale that sets the relevant mass/energy scale. (Corrections to the free 
Lagrangian of the form $v^4/r^n$ and $v^6/r^{2n}$, where $v$ denotes the velocity and $r$ distance, can be also taken into account in this calculation \cite{Liu:1997gk}.)
{\it Notice how this differs from the usual expression for the classical partition function which includes the factor 
$(1/N!)$.}\
This is a distinguishing feature of quantum Boltzmann statistics 
(as pointed out in \cite{Greenberg:1989ty, Greenberg:1991ec}).

Given the above expression for the partition function all other thermodynamic functions, including energy and entropy, are determined.
How can this expression be reconciled with non-extensive nature of entropy (holographic scaling) in the context of
gravity?
In order to obtain the holographic scaling for entropy we need a relation between $R$ and the size of the gravitational system, for example, a black hole.
The partition function of a free gas of distinguishable particles in a volume $V \sim b^{d-2}$ can be matched to the thermodynamic properties of a Schwarzschild black hole in $d$ spacetime dimensions
with the Schwarzschild radius $b$ \cite{Liu:1997gk}.
(Note that $d-2$ denotes the number of transverse dimensions, as required by the holographic scaling.)
As pointed out in \cite{Liu:1997gk}, in the limit in which the
thermal wavelength $\lambda = \sqrt{\beta R}$
of such a gas
is of the order of the Schwarzschild radius
$\lambda \sim b \sim V^{\frac{1}{d-2}}$,
the partition function $Z$, which can be
rewritten as 
\be
Z \sim \Big(\frac{V}{\lambda^{d-2}}\Big)^N,
\ee
and therefore, generically, $\log{Z} \sim N$, because $\lambda^{d-2} \sim V$.
(Note that this form of $Z$ follows from a direct computation of the canonical 
partition function for the case of infinite statistics,
without any reliance on holography and black hole physics, as discussed 
\cite{Goodison:1994aj}, and references
therein.)
In that limit, the temperature $T \sim \frac{1}{\beta}$ and the size $b$ match the formulae
for the temperature and the Schwarzschild radius,
respectively,
from black hole thermodynamics \cite{Liu:1997gk}.
In particular, the energy $E$ is calculated as
\cite{Liu:1997gk}
\be
E = - \frac{\partial \log{Z}}{\partial\beta } \sim \frac{N}{\beta},
\ee
and the entropy $S$ is given as \cite{Liu:1997gk}
\be
S = \log{Z} + \beta E \sim 2 N \sim N.
\ee
where the
numerical factor in front of $N$ is
not important for our central point for very large $N$ (as is the case in our context).
Thus, we obtain that the entropy is proportional to the
number of infinite statistics particles \cite{Liu:1997gk}
\be
S  \sim  N \to E \sim \frac{S}{\beta}.
\ee
As already pointed out, the requirement $\log{Z} \sim N$ amounts to the 
condition $\lambda \sim b \sim V^{\frac{1}{d-2}}$ \cite{Liu:1997gk}, or more explicitly
\be
 V (T/R)^{ (d-2)/2} \sim 1,
\ee
which, after using the interpretation of
the above energy $E$ as the light-cone energy related to the mass $M$ of a boosted object 
$E= \frac{R}{N} M^2$ \cite{Banks:1997hz}  (realized in Matrix theory, a light-cone formulation of M-theory) implies that the entropy scales as \cite{Liu:1997gk}
\be
S \sim (T/R)^{ - (d-2)/2},
\ee
and this is indeed true for Schwarzschild black holes and thus it
embodies the holographic principle
(in other words, the fact that
the black hole entropy scales with the area of the horizon). 
We also note that this argument can be
extended to the cosmological horizon, and the holographic scaling in that case,
which was relevant in our recent computation of the bound on the vacuum energy \cite{Freidel:2022ryr}.

The fluctuation of energy (the central quantity computed by Einstein in his seminal papers on early quantum theory and statistical physics) is another derivative 
with respect to $\beta$.
To be more precise given
\be
\langle E \rangle = \frac{1}{Z} \sum_i E_i e^{-\beta E_i},
\ee
then the fluctuation 
\be
\langle \epsilon^2 \rangle \equiv \langle E^2 \rangle - (\langle E \rangle)^2,
\ee
is determined as follows
\be
\langle \epsilon^2 \rangle = - \frac{\partial \langle E \rangle}{\partial \beta}.
\ee
From the expression for the energy we obtain that
\be
\langle \epsilon^2 \rangle \sim \frac{N}{\beta^2} \sim 
\frac{E}{\beta},
\ee
which then implies the 
result \cite{Verlinde:2019ade}
about the Brownian-motion-like geometric mean
\be
\sqrt{\langle \epsilon^2 \rangle} \sim 
\sqrt{\frac{E}{\beta}}.
\ee
The natural temperature $T \sim \frac{1}{\beta}$ for a gas of spacetime atoms
obeying infinite statistics, is by construction,
of the order of the Planck temperature (or equivalently, the Planck energy).
Thus, if we invoke the IR properties of quantum gravity, according to which higher energies $E$
and momenta correspond to larger distances $l$,
then we can write $E \sim l$, so that the
above equation becomes
$
\delta^2 \sim l\,l_P,
$
(another crucial equation from 
\cite{Verlinde:2019ade})
where $\delta$ denotes the fluctuation of length,
associated with the relevant fluctuation of energy.
We will comment on the phenomenological implications of this equation in the next subsection.

Note that this relation is true in any number of dimensions,
whereas, as discussed in the following subsection, our recent formula (reviewed in Section~\ref{s:CC}) for the vacuum energy scale is dimension dependent. (However, in 4 spacetime dimensions the two expressions coincide.) Also, this result is based solely on statistics and not on any particular model of quantum gravity.
At this point one might ask if given the above result one can go back and deduce that infinite statistics is inevitable.
Given what is known about the quon statistics \cite{Goodison:1994aj}, this statement is true: the only distinguishable statistics that supports the above relation between the fluctuation of energy and its average is indeed infinite statistics.
Also, given the available knowledge of the thermodynamics of infinite statistics
\cite{Goodison:1994aj}, it is clear that infinite statistics cannot be associated with any ``material'' (i.e., matter degrees of freedom), but that does not prevent it from being associated with the statistical properties of spacetime atoms.

This result is interesting because of the fact that modular cells are quantum objects and they are distinguishable and thus their statistics is quantum Boltzmann, or infinite statistics.
Thus if one models quantum spacetime as a free gas of objects 
obeying infinite statistics
(quanta of spacetime, with covariance being implied by the quantum nature of this description, as discussed in \SS\,\ref{s:QG=GQ})
then one
does obtain the result of \cite{Verlinde:2019ade} in any number of dimensions for this particular model of quantum spacetime.
(We emphasize a pleasing picture: the quantum matter degrees of freedom are fermionic, the 
quantum degrees of freedom of interaction forces are bosonic, and the quanta of spacetime obey infinite statistics.)

From this point of view one could propose the use of gravitational interferometry to probe this particular modular quantum ``spacetime foam'', and the above unique features of infinite statistics of modular cells (spacetime atoms).
This would be a clear analogy of Einstein's (and Smoluchowski's and Langevin's) theory of
Brownian motion and its experimental probe by Perrin \cite{Einstein:1956Inv, Perrin:1916Ato, Smoluchowski:1906Zur, Langevin:1908Sur}, however, in the domain of quantum gravity.
A concrete proposal is as follows:
Start with a theory formulated in modular spacetime,
such as the metastring.
Then we have natural matrix-like spacetime variables
(say for the metastring with a modular world sheet).
Consider a de~Sitter spacetime (with a characteristic scale --- the Hubble scale) as a bound state in such a theory (this is possible because the metastring theory is bosonic, and the metastring supports the computation of the vacuum energy reviewed in Section~\ref{s:CC}. de~Sitter spacetime has the Hawking-Gibbons temperature (de~Sitter temperature)
given by the inverse of the Hubble scale,
which is in turn related to the cosmological constant or vacuum energy of the metastring in that background.
The vacuum energy is, following the calculations from Section~\ref{s:CC}, bounded by the geometric mean of the Hubble and the Planck scales.

Then one should examine a Gibbs-like ensemble of matrix-like variables
at the de~Sitter temperature that describes the thermodynamics of de~Sitter bound state.
By construction such an ensemble of matrix-like variables obeys the rules of the Voiculescu 
non-commutative probability theory with infinite statistics.
Then the desired relation between the
variance of energy and its average (and the
square root number of degrees of freedom $N$) follows.
Note, that the modular Hamiltonian associated with
the de~Sitter horizon is the Hamiltonian of
the de~Sitter bound state --- so we also have the desired statement about the relation between the modular Hamiltonian and its variance\cite{Verlinde:2019ade}.

The spacetime “partons”, that is the elementary constituents, together with the background,
from which the de~Sitter bound state is made, should
be associated with matrix elements/entries, and the open strings between them.
Those are the “spacetime atoms” that obey
infinite statistics.
(As individual quanta, such as
the partons of Matrix theory, they are simply gravitons, with familiar spin-2 statistics. There is not contradiction here, because, as reviewed in Section~\ref{s:CC}, EFT with an extra input from holography leads to the same effective result for the vacuum energy as does a more fundamental description based on modular regularization of phase space and its consequences, including infinite statistics.)
Thus non-perturbative metastring theory provides a concrete model of modular spacetime and spacetime quanta obeying infinite statistics (consistent with the holographic bound), and having the free energy of an effective two dimensional theory, that leads to the relation between the fluctuation of the modular Hamiltonian and its expectation value, and furthermore, implies an effective collective field description of gravitons as spin-2 quanta of the effective field theory of the gravitational field, as well as the unique temporal dependence of the matrix model Green's function\cite{Gopakumar:1994iq} that could be experimentally tested.

\subsection{Quantum Spacetime and Quantum Gravity Phenomenology}
In this subsection we can pose the question of quantum gravity phenomenology associated with the generic
concept of modular spacetime and relate our discussion to
\cite{Verlinde:2019ade}
and the subject of experimental probes of quantum spacetime based on gravitational interferometry.

We start by repeating the argument we presented for the vacuum energy \cite{Freidel:2022ryr, Freidel:2023ytq}, as reviewed in Section~\ref{s:CC},
which was based on the modular regularization of
phase space and the holographic/Bekenstein bound,
albeit in the case of the causal diamond associated with
a gravitational wave detector (this set-up is observer dependent, and that fits our notion of relative (observer-dependent)
locality). In particular, we want to use Bekenstein's bound that is associated with the causal diamond of a gravitational interferometer.
Then we obtain that the characteristic scale of the vacuum (quantum spacetime) fluctuations is given by the see-saw (ie, geometric mean) formula
\be
\delta \sim \sqrt{l\,l_P}, 
\ee
where $l$ is a characteristic length associated with the interferometer, and $\delta $ is 
the characteristic length associated with the vacuum energy fluctuations due to the modular structure of covariant phase
space as in our recent papers  \cite{Freidel:2022ryr, Freidel:2023ytq}. As observed by Zurek, $\delta/l$ is of the order of the LIGO-VIRGO sensitivity ($\sim 10^{-20}$) for $l$ of order of a kilometer.
By our general argument, this could be a probe of modular spacetime and its modular structure, in analogy with the  
Brownian-motion-like probes of the atomic structure of matter \cite{Einstein:1956Inv, Perrin:1916Ato, Smoluchowski:1906Zur, Langevin:1908Sur}.
This reasoning makes sense, given Einstein's basic relation
$\langle x^2 \rangle \sim t$, between the average
of the distance squared and the elapsed time,
derived from the properties of the Gaussian distribution associated with the relevant diffusion equation of the Brownian movement.
In our case, $\delta^2 \sim \langle x^2 \rangle $,
and also $l \sim t$. 
The Gaussian distribution encountered in the classic analysis of Brownian motion is an
example of a classical Boltzmann distribution.
In our case we are dealing with quanta of spacetime that obey the quantum analog of the
Boltzmann distribution which leads to the same
relation between the fluctuation of energy and
its average, as derived in the previous subsection.

Note that in the case of the cosmological constant (cc) problem, $l$ was identified with the cosmological horizon (CH)
($l_{CH} \sim 10^{27}$\,m) which gave us the observed characteristic scale associated with the cosmological constant ($\delta_{cc} \sim 10^{-4}$\,m).
In the case of the gravitational wave interferometer, like LIGO, $l \sim 10^{3}$\,m and thus the characteristic vacuum 
energy scale in that case is $\delta \sim  10^{-16}$\,m. Also, the number of phase space cells in 4-dimensional spacetime was
$N_{cc} \sim l_{CH}^2/l_P^2 \sim 10^{124}$, whereas in the case of the gravitational interferometer $N \sim l^2/l_P^2 \sim 10^{76}$,
which implies that per direction of spacetime we have $N^{1/4} \sim 10^{19}$ ``spacetime atoms'' (which, as a historical remark, is not that different
from the value of the Avogadro number measured by Perrin following Einstein's, Smoluchowski's and Langevin's
theoretical work \cite{Einstein:1956Inv, Perrin:1916Ato, Smoluchowski:1906Zur, Langevin:1908Sur}).
In analogy with the classic treatment of Brownian motion \cite{Einstein:1956Inv, Perrin:1916Ato, Smoluchowski:1906Zur, Langevin:1908Sur} one should be able to measure this effective ``spacetime Avogadro number" $N^{1/4} \sim 10^{19}$ using
gravitational interferometry.

We note that the whole discussion presented in our recent paper on the vacuum energy problem and gravitational entropy that was specifically devoted to $4=3+1$ spacetime dimensions can be formally extended to
any $D{+}1$ spacetime dimensions.
We just summarize the relevant formulae (by setting all numerical factors to be of order one, for simplicity)\cite{Berglund:2022qsb}
\be
\begin{aligned}
   l^{D+1} \Lambda^{D+1} &\sim N,&\quad
  N &\sim \frac{l^{D-1}}{l_P^{D-1}},&\quad
  \Lambda^{D+1} &\sim \frac{1}{{l^2 l_P^{D-1}}},&\quad
  \lambda_{cc} &\sim \Lambda^{D+1} l_P^{D-1} \sim \frac{1}{l^2},
  & \quad
\end{aligned}\vspace*{-1mm}
 \label{e:4}
\ee
(The first formula follows from the volume of phase space, the second from the holographic bound in $D+1$ spacetime dimensions, the third expression is the consequence of the first two and the fourth expression is the definition of the cosmological constant in $D+1$ spacetime dimensions.)
Thus the geometric mean relation from 4 spacetime dimensions is generalized to $\Lambda \sim 1/\lL$
\be
\ \lL \sim l^{2/(D+1)} l_P^{(D-1)/(D+1)}
\ee
Indeed, for $D=3$ we recover the seesaw/geometric mean formula cited in the beginning of this subsection.

This general formula is {\it not} true for the situation considered in\cite{Verlinde:2019ade}: The geometric mean formula between the fluctuation of energy and the energy mean (as derived in the previous section) is
valid in any number of dimensions and it is a
square root/geometric mean/seesaw-like formula 
$\delta \sim \sqrt{l\,l_P}$ that we derived
using the general properties of infinite (or quantum Boltzmann, or quantum distinguishable) statistics.
However, in 4 spacetime dimensions the vacuum energy formula and the general formula derived from infinite
statistics regarding the relation between the energy fluctuation and its mean, coincide. Thus, gravitational interferometers could be used as probes of quantum distinguishable statistics associated with
4-dimensional spacetime quanta\footnote{Another important smoking gun for the modular spacetime is associated with
triple and higher order interference phenomena as discussed in the context of  the ``gravitization of quantum theory'' 
\cite{Berglund:2022qcc, Berglund:2022skk, Berglund:2023vrm}.}.

Moreover, as reviewed by Gross and Gopakumar\cite{Gopakumar:1994iq} (see also \cite{Douglas:1994zu}), the analog of the
Gaussian distribution for infinite statistics is the
Wigner semicircle law for the eigenvalues $m_i$ of an infinite
square matrix $\mathbb{M}$, so that the density $\rho(m)$ of eigenvalues
scales as
\be
\rho(m) = \frac{1}{2 \pi} \sqrt{4 - m^2},
\ee
for the Gaussian unitary ensemble
$\int D[\mathbb{M}] e^{-\frac{1}{2} \Tr[\mathbb{M}^2]} $.
This is in turn represented by the following operator expression in non-commutative probability theory
\be
\mathbb{M} = a + a^{\dagger},
\ee
with $a$ and $a^{\dagger}$ satisfying the free (Cuntz)
algebra $a a^{\dagger} =1$.
The above semicircle distribution for the eigenvalues of the
spacetime coordinate operators should be observed
in the experiments that probe infinite statistics
of spacetime quanta. This will be reflected in the time dependence of the relevant Green's function as implied by infinite statistics in the simplest case of the Wigner semicircle distribution, as discussed by Gross and Gopakumar\cite{Gopakumar:1994iq}.

\section{Quantum Gravity, UV/IR Mixing and Particle Physics} \label{s:PPP}
One of the main claims of this review is that quantum gravity
effects can be already detected in the context of the measurements of the cosmological constant, and that, viewed as gravitized quantum theory, quantum gravity can be probed in experiments that check for higher order (non-Born) quantum interference, as well as the probes of statistics of spacetime quanta using gravitational interferometry.
The main point here is that quantum gravity phenomenology occurs at scales far away from the Planck scale, and also in statistical, macroscopic settings. This is very much akin to the inherently quantum nature of macroscopic phenomena of black body radiation, or the hardness of matter, or other various
macroscopic effects such as magnetism, conductivity, etc.

Most of these macroscopic quantum effects are based on quantum (Bose-Einstein or Fermi-Dirac) statistics. As we argued in the previous section, in principle, spacetime quanta have infinite (quantum distinguishable) statistics and they can be probed via the gravitational analog of the Brownian motion experiment, but that probe is indirect. 
It is amusing to speculate that spacetime quanta might be confined in analogy with quarks, and this appears quite natural from the point of view of the matrix model realization. One reason for confinement of the spacetime quanta into what we identify as classical spacetime might be via the crucial appearance of higher order quantum correlations (responsible for higher order quantum interference) between these quanta. In analogy with other macroscopic quantum phenomena one could think about all sorts of properties of spacetime and their origins in quantum gravity: the origin of time\footnote{In string theory the Liouville field, coming from the conformal anomaly (a quantum effect), and representing the conformal mode of the world-sheet $1+1$ quantum gravity, appears as the natural clock, from the rewriting of the critical string theory as a non-critical string in one extra, timelike, dimension\cite{Polchinski:1998rq}.} from the rank and strength of the higher order correlations, pointing to the purely quantum origin of time; similarly, the quantum origin of space from the size of the higher order correlations; the origin of time arrow (and the impossibility of closed timelike curves) from the second law of thermodynamics that involves higher order correlations; the origin of causality (and thus, horizons) from the relation between the rank and the size of higher order correlations; the origin of inertial frames from the ``local'' quantum frames associated with ``local'' (in the space of states) quantum basis, and of inertia as the macroscopic leftover of the gravitization of quantum theory and the relation between short and long distances (or low rank and higher rank quantum correlations --- a sort of quantum Mach's principle); the equivalence principle as the left over of the QG=GQ dictionary (which could be understood as the quantum equivalence principle), etc. Of course, all these possibilities are quite speculative at the moment.

In this section we claim that the already observed spectrum of
masses of fundamental particles is indicative of quantum gravity phenomenology at much lower scales than the traditional Planck scale, $M_P\sim10^{19}$\,\GeV.
Following the presentation in the recent paper\cite{Berglund:2023gur}, we show that the above astro-particle conceptual arguments and ensuing calculations turn out to have rather concrete consequences regarding the --- {\em well measured} --- (Standard Model) particle physics. More precisely, we argue that the UV/IR mixing induces three separate ``intermediate'' mass-scales (see also Figure~\ref{f:Masses}):
 ({\small\bf1})~the Higgs mass-scale, $M_H\sim10^2\,\GeV$,
 ({\small\bf2})~the Bjorken-Zeldovich mass-scale, $M_{BZ}\sim7\,\MeV$,
 ({\small\bf3})~the mass-scale of the Standard Model decoupling from dark matter, $M_{SM}\sim4.5{\times}10^{14}\,\GeV$.
 In turn, these then induce the detailed pattern of precisely three generations of Standard Model fermion masses and their CKM/PMNS mixing. In particular, this enables us to write explicit predictions for the still undetected neutrino masses.

\subsection{Quantum Gravitational Roots of the Higgs Mass} 
\label{s:Higgs}
The underlying novelty that induces a stringy bound on the Higgs mass and the gauge hierarchy problem is the non-commutative chiral doubling of spacetime and its new Heisenberg algebra, $[x^\mu,\tilde x_\nu]=2\pi i\ls^2\,\delta^\mu_\nu$, etc., corresponding to~\eqref{xtxcomm}\cite{Freidel:2015pka, Freidel:2017xsi, Freidel:2017wst, Freidel:2017nhg}.
The non-commutativity of the dual coordinates is implied by a constant and nonzero Kalb-Ramond  $B_{\mu\nu}$-field (the axion in 4-dimensional spacetime), while dynamical backgrounds correspond to intrinsic non-associativity\cite{Freidel:2017nhg}.
Correspondingly, the zero modes in this metastring formulation of string theory are rigid length-$\ls$ strings that correlate each Standard Model particle with its dual,\footnote{Whereas the effective value of $\ls$ may well differ in 10 or any other spacetime dimension\cite{rBHM1,rBHM7}, here we focus on its 4-dimensional value.} the latter of which furnishing the most obvious candidate dark matter\cite{Berglund:2020qcu,Berglund:2021xlm}.

\paragraph{Cosmological Scale:}
Reapplied within the metastring framework and its modular phase space of the Heisenberg algebra~\eqref{xtxcomm} described in \SS\SS\,\ref{s:QSpT} and~\ref{s:mStr}, the above analysis of the vacuum energy (including the Bekenstein bound) now produces a bound on the Higgs mass and vev --- since {\em the latter specifies\/} the vacuum.

In a $4{+}4$-dimensional modular spacetime $(X^\mu,\Tilde{X}^\nu)$ with $N_{\Lambda}$ fluxes (see \SS\SS\,\ref{s:ModFlux}--\ref{s:mStr}) and respective length scales $\lL$ and $\Tilde{l}$,
\begin{equation}
    (l_\Lambda\, \Tilde{l}\,)^4 = N_{\Lambda}\, (\ls^2)^4
 \label{e:lLlN}
\end{equation}
is analogous to the relation~\eqref{e:lLN},
for the first of the Heisenberg algebras in~\eqref{xtxcomm}.
Together with the relation $N_{\Lambda}=\lL^2/l_P^2$, stemming from the holographic bound\cite{tHooft:1993dmi,Susskind:1994vu} for the effective spacetime associated with the vacuum energy, this produces
\begin{equation}
    \lL\,\Tilde{l}=\ls^2 \left(\frac{\lL^2}{l_P^2} \right)^{1/4} = 
    \ls^2 \left(\frac{\lL}{l_P} \right)^{1/2}.
\end{equation}
The string and the Planck lengths, $\ls,l_P$,  are related via the string coupling $g_s$
\be
g_s\,\ls = l_P, \quad\text{i.e.}\quad M_s = g_s\,M_P,
\label{e:Ms=gsMP}
\ee
with $M_s$ and $M_P$ the corresponding mass scales,
which sets the dual spacetime scale:
\be
    \tilde {l} = \frac{\ls^2}{\sqrt{\lL\, l_P}}
    = \frac{l_P}{g_s^2} \Big(\frac{l_P}{\lL}\Big)^{\!1/2}.
\ee 
Writing $\tilde{l}=\eta\,l_P$ produces (see Table~\ref{t:Masses}):
\be
g_s^2=\frac1\eta\sqrt{\frac{l_P}{\lL}}=\frac1\eta\sqrt{\frac{M_\L}{M_P}}
\quad\Longrightarrow\quad
g_s<1~~\text{when}~~\eta\define \frac{\tilde{l}}{l_P}
\gtrsim 10^{-15.5}.
\label{e:gs=small}
\ee
That is, $g_s<1$ and the one-loop computation of the (metastring) partition function is a good approximation as long as $\tilde{l}\lesssim10^{-21}$\,m, and certainly so if the ``dual (part of) phase space'' (defined by the dual spacetime and dual momenta) is of Planck size --- which we assume hereafter, for simplicity.

\paragraph{Higgs Mass:} A formula for the Higgs mass was recently obtained (in bosonic string theory) by Abel and Dienes\cite{Abel:2021tyt},
\be
M_H^2 = \xi\frac{M_\L^4}{M_P^2} - \frac{g_s^2 M_s^2}{8\pi^2} \vev\cX,
\label{e:AbelDienes}
\ee
by relying primarily on modular invariance, which is however a universal stringy feature. The $\xi$-term provides the modular completion to the
second term, with a suitably normalized insertion in the second moment of the partition function, $\cX$, for which $\vev\cX={-}|\vev\cX|$.
 This result in fact also follows from the foregoing analysis in the metastring framework: Indeed, using~\eqref{e:Ms=gsMP} and~\eqref{e:gs=small} recasts~\eqref{e:AbelDienes} as
\begin{equation}
  M_H^2
  =\xi\Big(\frac{M_\L^2}{M_P}\Big)^2
          +\frac{|\vev\cX|}{8\pi^2}\big(\sqrt{M_\L M_P}\big)^2,
\label{e:AbelDienes2}
\end{equation}
which is a simple numerical combination\footnote{$|\vev\cX| \sim 10^{-1}$ is consistent with Abel \& Dienes' results\cite{Abel:2021tyt}.} of the
 familiar ``seesaw'' ($M_\L^2/M_P$) and
 geometric mean ($\sqrt{\,M_\L M_P}$) terms,
both reflecting the seesaw relation of two scales, $M_\L$ and $M_P$.

The second term dominating the first one ($\sim10^{-34}$\,eV), we have
\be
M_H \sim g_s\, M_s \sqrt{\frac{|\vev\cX|}{8 \pi^2} }
 \,=\, g_s^2\, M_P \sqrt{\frac{|\vev\cX|}{8 \pi^2} }
 \,\sim\, \sqrt{M_\L M_P }  \sqrt{\frac{|\vev\cX|}{8 \pi^2} },
\label{e:mH-BHM}
\ee
recovering the string-theoretic seesaw formula for the Higgs mass\cite{Berglund:2022qsb}, with the very realistic numerical value as shown in Table~\ref{t:Masses}.
We emphasize that~\eqref{e:mH-BHM} is to be understood as a bound on the Higgs mass, as is~\eqref{e:lcc=llP} for the cosmological constant.

\paragraph{Summary:}
Akin to our cosmological constant arguments and result~\eqref{e:lcc=llP}, a stringy seesaw formula~\eqref{e:mH-BHM} also follows for the Higgs mass:
 ({\small\bf1})~within the metastring formulation of string theory and its modular spacetime, by
 ({\small\bf2})~combining the $[x,\tilde{x}]\neq0$ non-commutativity and holography in $x$-space,
 and by
 ({\small\bf3})~assuming that $\Tilde{x}$ is of the Planck length size. 
As with the vacuum energy, \eqref{e:mH-BHM} is also a bound provided by the size of the phase space and the Bekenstein bound in which the effective length scale is associated with vacuum energy $\lL$.
In this calculation, the two Heisenberg algebras in the metastring formulation ($[x,p]$ and $[x,\tilde{x}]$) are mutually consistent.

Having applied this logic to the formula for the Higgs mass \`a la Abel \& Dienes\cite{Abel:2021tyt} (derived in canonical bosonic string theory with manifest stringy modular invariance, but also compatible with its metastring formulation), 
we have not only arrived at a stringy bound for the Higgs mass, but also at a completely stringy view of the hierarchy problem. So, both the Higgs mass and the hierarchy problem are direct consequences of the new view of quantum gravity as gravitized quantum theory. The hierarchy problem is directly tied to the vacuum energy problem, whereby the resolution of both lies in the fundamental (modular) phase-space approach combined with a Bekenstein bound on the number of relevant degrees of freedom.
This new and unified understanding of these two central hierarchy problems
naturally points to metastring theory, and (as we outline in the next section) it can also address the problem of fermion masses.

Let us however conclude this section by addressing the naturalness of the above values for $N$, relevant for {\em\/both\/} hierarchy problems: the cosmological constant and the Higgs mass.
Both in statistical physics and in QFT, it is well known how to sum over contributions of closed diagrams:
 Simple combinatorics ensures that this is an exponent of the partition function associated with a closed loop (handle, for strings).
 As pointed out in Section~\ref{s:CCP}, the QFT vacuum partition function is 
 $Z_{\text{vac}} = \exp(Z_{S^1})$, with $S^1$ the circle of a vacuum loop traced by a particle;
 in string theory, one just replaces $S^1\to T^2$\cite{Freidel:2022ryr}.
 For the case of {\em dynamical} Born geometry\cite{Freidel:2014qna} (a generic feature of quantum gravity in the metastring formulation), the usual path integral measure $e^{iS}$ should be effectively replaced  by $e^{e^{iS}}$ after summing over handles of a dynamical quantum geometry\footnote{It is tempting to associate this double exponential with another severe fine tuning problem in cosmology --- that of the initial state\cite{Berglund:2022qsb}.}, where in the approximation of a dilute gas of handles we have taken that the
effective partition function is just the canonical one.
 Summing over handles in the foamy quantum space\cite{Berglund:2023vrm}, from the point of view of the canonical complex geometry of quantum theory, thereby yields an effective action which is essentially $e^{iS}$.
 In the Euclidean formulation, this implies that the effective action at some scale sensitive to gravity can be {\em exponentially removed} from the natural scale of Planck gravity, indicating that the Higgs scale may well be where effects of quantum gravity could be seen.\footnote{Indeed, see the large class of widely usable toy models\cite{rBHM1,rBHM7,Berglund:2020qcu,Berglund:2021xlm}.}
Essentially, we claim the naturalness of the hierarchy of scales between the Higgs and
the Planck scale ultimately to be a quantum gravity effect, associated with ``gravitizing the quantum''\cite{Freidel:2014qna,Berglund:2022qcc,Berglund:2022skk}.
 Thus, the effective value of $N$ (per spacetime direction) that features in both hierarchy problems, the cosmological constant problem and the problem of the Higgs mass, is indeed naturally expected to be of the order of the familiar Avogadro number, and it is only genuine in the context of quantum gravity (or gravitized quantum theory) and quantized spacetime.

\subsection{Quantum Gravity and Fermion Masses: General Comments}
\label{s:Generalia}
The foregoing discussion shows that the same reasoning and computation successfully produces seesaw-like formulae,
 \eqref{e:lcc=llP} for the cosmological constant, and
 \eqref{e:mH-BHM} for the Higgs mass.
The weak string-coupling estimate~\eqref{e:gs=small} is consistent with the lowest-order perturbative computations and analysis of the string partition function, and supports such expectations also for the (meta)particle limit.
Indeed, for both particles and strings, the cosmological constant was thus shown to be bounded by the phase space volume, its modular regularization and Bekenstein bound.
In turn, Abel and Dienes' stringy Higgs mass formula~\eqref{e:AbelDienes2}\cite{Abel:2021tyt} produces the above, evidently analogous result --- the logic of which is not attainable in EFT.

Both of these results follow from the new, ``QG=GQ'' view of quantum gravity as gravitized quantum theory.
The analogous solutions to these hallmark problems only have differing contextual choices of the UV and IR scales:
For both, $M_P$ is the natural UV scale. On the other end,
the Hubble/cosmological horizon provides a natural IR scale ($M_{CH}$) for the cosmological constant scale, $M_\L$~\eqref{e:lcc=llP}, which in turns is the IR scale for he Higgs mass, $M_H$ in~\eqref{e:mH-BHM}.
The result of this reasoning is bound to agree with Abel and Dienes's explicit stringy result~\eqref{e:AbelDienes2} since the vev of the Higgs field {\em\/specifies\/} both the Standard Model vacuum as well as the Higgs mass.

In fact, since the same Higgs vev also provides masses to all charged Standard Model fermions, the above reasoning should extend also to those. Whereas $M_H$ is also relevant for neutrinos, the above reasoning will extend to them differently. However, before delving into a derivation of the cascading seesaw formulae in Table~\ref{t:Masses}, several frame-setting comments and observations are in order.

\paragraph{Criticality:}
whereby the top-quark mass may be related to the Higgs mass as proposed by Froggatt and Nielsen\cite{Froggatt:1995rt}, is our first motivation. The running top and the Higgs mass are related through the running top Yukawa coupling and the Higgs self coupling evaluated at the running scale $\mu$ that is given by the top and Higgs masses. Given the top pole mass $m_t$, that is the physical mass of the top, and given the top running mass $M_t(\mu)$\cite{Froggatt:1995rt}
\be
\frac{m_t}{M_t(m_t)} = 1 + \frac{4}{3} \frac{\alpha_s(m_t)}{\pi} +
10.95 \big(\frac{\alpha_s(m_t)}{\pi}\big)^2,
\ee
where $\alpha_s(m_t)$ is the running strong coupling constant evaluated at the top mass.
The Higgs mass is already determined in the previous section, and we
only have to set the running scale $\mu$ at the first and the second minimum of the
Higgs potential, which by criticality, are given by the electroweak scale and the Planck scale, respectively\cite{Froggatt:1995rt}. The mass of the top follows then from
the mass of the Higgs. (One could repeat the same analysis for the bottom quark and the tau lepton.)
This, in turn, implies, via~\eqref{e:AbelDienes2}, that the mass of the top also could be related to
the cosmological constant --- because the Higgs mass is. 
Again by dimensional analysis, as in~\eqref{e:AbelDienes2}--\eqref{e:mH-BHM}, the analogous fermionic formulae are expected to be of the form $m_\psi \sim g_s M_s$,
up to the multiplicative coefficients implied by stringy modular invariance. 
This suggests a seesaw formula akin to the one for the Higgs mass~\eqref{e:mH-BHM},
however with appropriate UV- and IR-scales.\footnote{This indeed follows Weinberg's general idea, ``in some leading approximation the only quarks and leptons with nonzero mass are those of the third generation, the tau, top, and bottom, with the other lepton and quark masses arising from some sort of radiative correction''\cite{Weinberg:2020zba} --- except, the lower fermion masses are here generated by variants of the T-duality seesaw mechanism from a stringy non-perturbative effect; see \SS\,\ref{s:mStr}.} The claim here is that such seesaw formulae relate seemingly independent fermionic masses (in different generations) in the Standard Model.
In essence, this reasoning provides for the origin of different generations, starting from the heaviest fermions, and predicts that there can exist no heavier generations of Standard Model fermions.

To assess appropriate UV and IR scales for such fermionic seesaw formulae,
recall that the charged fermion masses ($m_t$, $m_b$ and $m_\t$) are related via the RG equations for the heaviest fermions in explicitly computable stringy models\cite{Faraggi:1991be}, and thus are natural candidates for the UV scales.
As to an appropriate IR scale, we present below an entropy argument that leads
to a scale attributed to QCD, but an order of magnitude smaller than the standard $\L_{QCD}$; we call this the Bjorken-Zeldovich scale, $M_{BZ}\simeq7$\,MeV.
This new low energy scale (that is very close to the scale of Big Bang Nucleosynthesis) can be also viewed as another fundamental infrared quantum gravity scale.
We then find (as observed by Bjorken in a completely different context\cite{Bjorken:2013aa}) that this $M_{BZ}$ can, with the masses of the heaviest charged fermions as the UV scale, parametrize the masses of all remaining charged fermions.

This observation can be properly justified only by
a computation of the bound of the partition function
of the Standard Model in the modular polarization, which by the already explicit computation of the cosmological constant is given by the volume of phase space. 
Relating then the number of phase space cells, in modular regularization, to the Bekenstein-like bound
with the UV scale given by the masses of the heaviest quarks and the heaviest lepton then reproduces Bjorken's expressions\cite{Bjorken:2013aa,rBJ-MM}.

\paragraph{Seesaw Structure:}
The foregoing discussion, including the stringy result~\eqref{e:AbelDienes2}, involves two types of formulae:
The geometric mean, $m<(m'\,{\sim}\,\sqrt{mM})<M$, is here implied by the non-commutative, symplectic structure of Born geometry, $\omega_{ab}$ in~\eqref{etaH0}.
The standard ``seesaw,'' $(m''\sim m^2/M)<m<M$, is familiar from neutrino physics and is here of the T-duality type, implied by the bi-orthogonal structure of Born geometry, $\eta_{ab}$ in~\eqref{etaH0}. The presence of the double metric, $H_{ab}$ in~\eqref{etaH0}, is what allows the doubling of the heaviest mass in the first place.  This provides for three distinct masses and is, essentially, our key observation here.

This dovetails with the fact that there are three generations, and
meshes nicely with the present experimental constraints on
other generations of quarks and leptons. 
In what follows, the bounds on the charged fermion masses
take the form of these seesaw relations
(as used for the cosmological constant and also for the Higgs mass):
with $M_{UV}$ identified with the heaviest mass,
the lighter copies are $M_{IR}$-multiples of numerical factors that are solely the square-root of ratios of the UV and IR scales, or the other way around:
\begin{subequations}
 \label{e:twoRoots}
 \begin{alignat}9
    M_{IR} \sqrt{\frac{M_{UV}}{M_{IR}}}&=\sqrt{M_{IR}M_{UV}},
     &\qquad&\text{ for the middle, and} \label{e:twoRoots1}\\
    M_{IR} \sqrt{\frac{M_{IR}}{M_{UV}}}
     &=\sqrt{\Big(\frac{M_{IR}^2}{M_{UV}}\Big)\,M_{IR}},
     &\qquad& \text{for the lightest.}\label{e:twoRoots2}
 \end{alignat}
\end{subequations}
With the UV and IR scales as reasoned above, one expects the numerical factors in~\eqref{e:twoRoots} to be square-roots of their ratios. Analogously, the dominant $\cX$-term in~\eqref{e:AbelDienes2} gives $M_H\sim g_s^2M_P=\sqrt{M_\L/M_P}\,M_P=M_\L\sqrt{M_P/M_\L}$. Higher powers of these square-root factors then correspond to higher powers of $g_s^2$, and are expected as (string-perturbative) {\em corrections\/} to~\eqref{e:AbelDienes2}.\footnote{Also, the evident $g_s\to g_s^{-1}$ map between~\eqref{e:twoRoots1} and~\eqref{e:twoRoots2} would seem to indicate that S-duality must be involved in an underlying stringy derivation of such formulae.} By the same token, higher powers of the square-root factors in~\eqref{e:twoRoots} are expected as corrections of these formulae.
 For example, the {\em standard} seesaw-formula from the original, neutrino physics,
\begin{equation}
  \frac{M_{IR}^2}{M_{UV}} = M_{IR}\,\Big(\frac{M_{IR}}{M_{UV}}\Big)
  = M_{IR}\,\Big(\sqrt{\frac{M_{IR}}{M_{UV}}}\Big)^2,
 \label{e:noRoots}
\end{equation}
features the {\em\/square\/} of the numerical factor in~\eqref{e:twoRoots2}, and is expected to correspond to an {\em additive} correction to~\eqref{e:twoRoots2}.

Also, let's assume that a fermionic version of the stringy result~\eqref{e:AbelDienes} can be derived, with a corresponding insertion vev, $\vev{\cX_\j}$, proportional to the gauge charges of the fermion $\j$ as indeed {\em is\/} the case for the Higgs field\cite{Abel:2021tyt}.
Then:
 ({\small\bf1})~for charged leptons, $\cX_\j\neq0$, the second term in a \eqref{e:AbelDienes}-like formula again dominates, and formula~\eqref{e:twoRoots1} follows.
 ({\small\bf2})~For chargeless neutrinos, $\cX_\j=0$, only the first term in a \eqref{e:AbelDienes}-like formula remains, and~\eqref{e:noRoots} is the {\em only} contribution.

The remaining (T-duality type) seesaw formula~\eqref{e:twoRoots2} 
stems from the central property of the zero modes of the metastring
captured by the action of the metaparticle~\eqref{mp1}, and 
especially the constraint 
$p{\cdot}\tilde{p} = \mu$.
This is precisely the second, ``seesaw-light'' type relation, where we identify $\mu=M_{BZ}^2$ and the size of the dual momentum space with the relevant charged fermion mass.
Unlike the first seesaw formula~\eqref{e:twoRoots1}, which essentially follows from the phase-space-like structure and so is associated with the symplectic form, this second seesaw formula~\eqref{e:twoRoots2} is induced by the bi-orthogonal structure of Born geometry.

Ideally, one would need a precise fermionic analogue of the Abel-Dienes formula for the Higgs mass in string theory\cite{Abel:2021tyt}.
In the absence of such explicit formulae, we identify key seesaw features that connect our approach to  Bjorken's observations\cite{Bjorken:2013aa,rBJ-MM}, which we then also extend to the CKM matrix (like Bjorken), but also to neutrinos and the PMNS matrix (in ways different from Bjorken).
We find it intriguing that
the same logic used for the computation of the cosmological constant extends, first to the Higgs mass,
and then also to the masses of all quarks and leptons. In fact, this strongly suggests and a precise fermionic analogue of the Abel-Dienes formula\cite{Abel:2021tyt} must exist.

While these seesaw features  do appear to be  cohesive and coherent,
a firm proof  would require the formulation of an explicit treatment of the Standard Model (SM) as a {\em\/modular QFT\/}: Every SM field $\phi$ is defined over both spacetime and the dual (momentum-like) spacetime,
$\phi(x, \tilde{x})$, with an intrinsic non-commutativity\cite{Freidel:2017xsi,Freidel:2018apz}, 
$[x, \tilde{x}] = i \ell_{nc}^2$, where $\ell_{nc}$ is in principle contextual, and not necessarily the string length or the Planck length.\footnote{Both of these scales, $\ls$ and $l_P$, turned up naturally in the discussion in Section~\ref{s:mStr}, but note that the effective, physically relevant 4-dimensional Planck scale may be removed, even exponentially much, from the underlying fundamental scale, e.g., in the large class of models discussed in\cite{rBHM1,rBHM7,Berglund:2020qcu,Berglund:2021xlm}.} By construction, such a formulation would have a natural solution of the vacuum energy problem,
and then, we conjecture, would also lead to the formulae for the fermionic masses presented below.
Such a modular SM would thereby imply relations between masses of different fermion generations that are invisible to the standard QFT form of the SM.
Such a modular QFT form of the SM can be also embedded in the metastring, which suggest a completely new (and complementary)
view on the origin of the SM in string theory, as compared to the traditional one based on Calabi-Yau compactifications in the point-field limit QFT\cite{Polchinski:1998rq}.
This should indicate that there are missing concepts (modular spacetime, modular polarization, Born geometry,
modular fields, metaparticles and metastrings) in the usual approach, and that the introduction of these missing
concepts to the canonical approach would yield the results discussed in this paper.

Unlike the very concrete foregoing statements about the vacuum energy problem and the problem of the Higgs mass, our present discussion of fermion masses is just a working conjecture at the moment.
We now turn to the implementation of this general set-up by
following our recent presentation\cite{Berglund:2023gur, Minic:2023oty}.

\subsection{Quantum Gravity and Three Generations of Fermion Masses}
\label{s:fMasses}

\paragraph{The Bjorken-Zeldovich scale:}
The relation~\eqref{e:NllP} specified the ``spacetime Avogadro number,''
 $N^{1/4} \sim \sqrt{l/l_P}\sim10^{31}$. However, in 3-dimensional space and as expected from {\em\/extensive\/} non-gravitational entropy, $N$ {\em\/defines\/} a length-scale, $l_{BZ}$ (see below for naming):
\be
l^3/l_{BZ}^3 \overset{\sss\text{def}}\sim N \sim  l^2/l_P^2,
\quad\Rightarrow\quad
l_{BZ}^3 \overset{\sss\text{def}}\sim l^3/N \sim l\,l_P^2 
\overset{\sss\eqref{e:lcc=llP}}{\sim} l_{cc}^2\,l_P,
~~~\text{i.e.},~~~
M_{BZ}^3 \sim M_\L^2\,M_P.
\label{e:lBZ}
\ee
More precisely, the $l_{BZ}$ and $M_{BZ}$ scales have been {\em\/deduced\/} from:
 ({\small\bf1})~our $N$~\eqref{e:NllP},
 ({\small\bf2})~the Bekenstein bound for gravitational degrees of freedom,
 ({\small\bf3})~the fact that matter and spacetime degrees of freedom are ``two sides of the same coin'' in (meta)string theory, and
 ({\small\bf4})~the extensive nature of entropy for the matter degrees of freedom.

The numerical estimate in Table~\ref{t:Masses}, $M_{BZ}\simeq 7.2\,\MeV$, turns out to be {\em\/exactly\/} the value used by Bjorken \cite{Bjorken:2013aa}\footnote{Bjorken seems to have been inspired by the work of the Oxford group\cite{Chan:2015bvx}, and discussed $M_{BZ}$ in a radically different context of the MacDowell–Mansouri approach to gravity, and in particular, the Friedmann-Robertson-Walker cosmology in that formulation\cite{Bjorken:2013aa}.} to parametrize the observed masses of Standard Model quarks and leptons, which is why we call $M_{BZ}$ the ``Bjorken-Zeldovich scale.''
The continued relations~\eqref{e:lBZ} express it in terms of the cosmological horizon and the Planck scale, and provides a derivation that is, as best as we know, completely new.
These three key scales,
the cosmological constant mass scale, $M_\L$, 
the Higgs mass scale, $M_H$, as well as 
the Bjorken-Zeldovich scale, $M_{BZ}$,
are thereby all ultimately determined in terms of the Hubble ($M_{CH}$) and Planck mass ($M_P$) scales; see Table~\ref{t:Masses} and Figure~\ref{f:Masses}.

There is an another, exceedingly curious relation, and independently corroborating the preceding derivation: The seesaw expression with the proton mass $m_p$,
\begin{equation}
  \frac{{m_p}^{\,2}}{M_H}=\frac{(938.27\,\MeV)^2}{125.25\,\GeV}
  \simeq7.0288\,\MeV
 \label{e:mp2/mH}
\end{equation}
{\em almost exactly\,} reproduces the above-derived $m_{BZ}$~\eqref{e:lBZ}! Furthermore, the structure of the seesaw formula is itself aligned with our observations throughout this report: 
The well understood electroweak phase transition (i.e., the Higgs mass, derived in \SS\,\ref{s:Higgs}, above) sets the UV scale (in the denominator) of~\eqref{e:mp2/mH}. In turn, the numerator specifies an appropriate IR mass-scale, well known to be associated with QCD: it is the mass of the {\em lightest and only stable QCD bound state,} the proton.\footnote{The commonly cited Landau pole, $\L_{QCD}\sim100\,\MeV$, is the momentum exchange magnitude where the {\em\/perturbatively\/} computed coupling parameter diverges, and so seems to be a hallmark of {\em\/a perturbative description\/} rather than an intrinsic characteristic of QCD itself. Also, likeliest candidates for pure-glue bound states have masses above 2\,\GeV\cite{Ablikim:2024Det}.} This then identifies the mass-scale~\eqref{e:mp2/mH}, equal to $M_{BZ}$, as characteristic for {\em\/charged\/} SM-fermions: The QCD-characteristic~\eqref{e:mp2/mH} is evidently relevant for quarks, and then also to electrically charged leptons, via EM-radiative corrections. Analogous radiative corrections to neutrino masses are below $M_H$ suppressed by weak gauge boson masses, and will have to be determined by different mass-scales; see~\eqref{e:m3}--\eqref{e:m1}, below.
Let us also add that the oft-cited and ${\sim}\,20$ times larger $\L_{QCD}\sim150\,\MeV$ is defined by the Landau pole, i.e., the momentum transfer at which the ({\em\/perturbatively computed\/}) strong interaction coupling diverges; $\L_{QCD}$ is thus a characteristic of the perturbative description of QCD, rather than of QCD itself.

\paragraph{Quarks:}
Following the above reasoning, we start with the masses of the heaviest charged SM fermions, $m_t, m_b, m_\tau$, as essentially being determined by the electroweak phrase transition, i.e., the Higgs mass. To this end, we write (see Table~\ref{t:Masses})
\begin{alignat}9
  m_t        &\define \2{Y_t}M_H,&\quad
  m_b        &\define \2{Y_b}M_H,&\quad&\text{and}&\quad
  m_\tau     &\define \2{Y_\tau}M_H,\\
\text{were}\qquad
  \2{Y_t}    &\approx \,^7\!/\mkern-1mu_5,&\quad
  \2{Y_b}    &\approx \,^{\2{Y_t}}\!/\mkern-1mu_{42},&\quad&\text{and}&\quad
  \2{Y_\tau} &\approx \,^{\2{Y_t}}\!/\mkern-1mu_{100},
\end{alignat}
are the concrete numerical values of these Yukawa couplings; see below for a justification of these estimates.
(For a concrete computation of these masses in a string theory model, see \cite{Faraggi:1991be}.) These masses serve as analogs of the UV scale in our Higgs mass formula~\eqref{e:mH-BHM}, whereas the
Bjorken-Zeldovich scale, $M_{BZ}\approx7$\,MeV, acts as the natural IR scale.
Note that the top mass $m_t$ is essentially tied to the Higgs
scale,\footnote{To this end, we cite the well-known argument based on criticality of the Standard Model that relates the masses of the top quark and the Higgs boson
\cite{Froggatt:1995rt} (see also \cite{Donoghue:2005cf,Khoury:2022ish}, for landscape-motivated discussions).} which in turn is
given by the (geometric mean) seesaw formula of the vacuum energy scale and the Planck scale.
Thereby, the top quark mass is ultimately also given in terms of the Hubble and Planck mass scales.
Analogously to~\eqref{e:mH-BHM} for the Higgs mass, 
the (geometric mean) seesaw relation then produces the charm mass
in terms of $M_{BZ}$ and $m_t$ (cf.\ the observed value in parentheses\cite{rPDG22}):
\be
m_c \sim \sqrt{M_{BZ}\, m_t} = M_{BZ} \sqrt{\frac{m_t}{M_{BZ}}}
\sim 1.10~(1.27)\,\text{GeV}.
 \label{e:mch}
\ee
Next, using the bottom-quark mass scale\footnote{For example, explicit calculation in the stringy calculation\cite{Faraggi:1991be} ties, via RG equations, the mass of the top quark to the mass of the bottom quark and the tau lepton, and so are all ultimately determined by the Hubble and Planck mass scales.} 
(instead of $m_t$)
and the same Bjorken-Zeldovich scale as the characteristic vacuum energy scale of matter,
the same seesaw relation yields the mass of the strange quark
\be
m_s \sim \sqrt{M_{BZ}\, m_b} =
M_{BZ} \sqrt{\frac{m_b}{M_{BZ}}}
\sim 171~(93.4)\,\text{MeV}.
 \label{e:mst}
\ee
Bjorken estimates 
the up- and down-quark masses essentially at
the Bjorken-Zeldovich scale:
$m_u \sim M_{BZ}$ and  $m_d \sim M_{BZ}$,
but models the actual relation $m_d > m_u$ with {\em ad hoc\/} factors~\cite{Bjorken:2013aa}.
Independently, the masses of the lightest quarks may be
deduced from chiral perturbation theory as
$m_u \sim 2$\,MeV, $m_d \sim 5$\,MeV.
However, apart from non-commutativity that led to~\eqref{e:mch} and~\eqref{e:mst}, 
our seesaw structure reasoning above involves 
also the inherent metastring/metaparticle T-duality, 
which induces the familiar ``seesaw-light'' relation.
This then leads to the following estimates (actual values in parentheses\cite{rPDG22})
\be
m_u \sim M_{BZ}^2/m_c \sim
M_{BZ} \sqrt{\frac{M_{BZ}}{m_t}}
\sim 10^{-2}M_{BZ} \sim 10^{-1}~(2.16)\,\text{MeV}. \label{e:mup}
\ee
This estimate turns out too small (by a factor of about 50),
but is (importantly!) smaller than the down quark mass estimate (also too small by a factor of about 16), 
\be
m_d \sim M_{BZ}^2/m_s \sim
M_{BZ} \sqrt{\frac{M_{BZ}}{m_b}}
\sim 10^{-1} M_{BZ} \sim 1~(4.67)\,\text{MeV}.  \label{e:mdn}
\ee
The above reasoning thus automatically reproduces the 1st generation ``mass inversion'':
 $\eqref{e:mch}>\eqref{e:mst}$ but $\eqref{e:mup}<\eqref{e:mdn}$, which is necessary for the proton to be stable while the neutron decays.
Thus, given the heaviest, top and the bottom quark masses,
the two distinct seesaw type formulae (non-commutativity and T-duality) produce quite realistic estimates for the masses of the middle and the lightest quark generations.

\paragraph{Charged leptons:}
Turning to the charged leptons, the evident analogue of the top-quark is the tau-lepton.
From a naive stability analysis of the tau analogue of the hydrogen atom,
the mass of the tau is expected to be of the order of 
the mass of the nucleus, i.e. a GeV.
This is supported since the masses of the top, bottom quark and the tau lepton are all related by the RG equations, as in the calculation of\cite{Faraggi:1991be}.
With the tau mass as given (again, from the calculation of\cite{Faraggi:1991be}, and ultimately related to the Hubble and Planck mass scales, much as the top and bottom quark masses are), the (geometric mean) seesaw estimate of the muon mass is (actual value in parentheses\cite{rPDG22}):
\be
m_{\mu} \sim \sqrt{M_{BZ}\, m_{\tau}} =
M_{BZ} \sqrt{\frac{m_{\tau}}{M_{BZ}}}
\sim 112~(106)\,\text{MeV}.
\ee
Just as with quarks, the second (T-duality kind) seesaw relation then yields the electron mass, given the calculated muon mass
\be
m_e \sim \frac{M_{BZ}^2} {m_{\mu}} \sim
M_{BZ} \sqrt{\frac{M_{BZ}}{m_{\tau}}}
\sim 464~(511)\,\text{keV}.
\ee

This proposal thus reproduces 3 generations of charged Standard Model fermions and their masses,
by the framework of the dual space, the modular spacetime Born geometry, and ultimately the metastring, i.e., by the intrinsic non-commutativity and covariant T-duality of the metastring.
The masses of the two lighter generations are induced from the masses of 
the heaviest quarks and leptons, and 
are fixed by non-commutativity and T-duality,
in analogy with the reasoning that gives the Higgs mass and the cosmological constant.
All of these formulae are seesaw-like and contextual bounds.
All of them ultimately reduce to the
IR size of the universe and the UV Planck length.

\paragraph{Neutrinos:}
Turning to neutrino masses and following Weinberg's original dimension-5 operator proposal in the Standard Model\cite{Weinberg:1979sa} (implying Majorana masses as well),
we estimate the heaviest (``tau'') neutrino mass to be
\be
m_3 \sim M_H^2/M_{SM} \sim (10^{-1} - 10^{-2})\,\text{eV}, \label{e:m3}
\ee
where the SM scale $M_{SM}$ is given by a  ``would-be unification scale'' of
the SM couplings (as indicated by RG equations), $\sim10^{15-16}$\,GeV, 
and $M_H$ is the Higgs scale of around $1$\,TeV.
This Standard Model scale can be also related to the Hubble and the Planck scales $M_{SM} = M_{CH}^{1/14} M_P^{13/14}
\sim4.5{\times}10^{14}\,\GeV$, as indicated in Table~\ref{t:Masses}. This $M_{SM}$ also appears in Vafa's analysis\cite{Vafa:2024fpx}, postulated as the scale at which decaying dark matter ``gives a small kick to its decay products.'' The fully T-dual, chirally doubled description of stringy spacetime\cite{Freidel:2015pka}, dark and visible matter mix via a Berry phase-like correlation term\cite{Berglund:2020qcu}. The mass-scale of this correlation term defines the decoupling energy, below which the dark and visible matter interact as required, only gravitationally. $M_{SM}\simeq4.5{\times}10^{14}\,\GeV$ is thus naturally identifiable with this dark-matter/visible-matter decoupling scale.

The mass of the heaviest neutrino~\eqref{e:m3} would then also be ultimately given 
in terms of the Hubble and the Planck mass scales.
The middle (``muon'') neutrino mass is then given
by a (geometric mean) seesaw formula, involving a low vacuum energy scale.
Unlike all quarks and charged leptons, neutrinos do not get
their masses from the Higgs mechanism, so the vacuum scale
cannot be $M_{BZ}$ (characteristic for quarks and charged fermions) and so must be 
the only other vacuum scale:
the cosmological vacuum scale associated with the cosmological constant~\eqref{e:lcc=llP}:
\be
m_2 \sim \sqrt{M_\L\, m_3} =
M_\L \sqrt{\frac{m_3}{M_\L}}
\sim (10^{-2} -10^{-2.5})\,\text{eV}. \label{e:m2}
\ee
By comparison, a similar mass value has been argued\cite{Aydemir:2017hyf}
to be natural by examining a dimension 6 analogue of Weinberg's operator, where a neutrino 
could acquires its mass from a fermionic condensate controlled by the
Bjorken-Zeldovich scale, with the electroweak cutoff scale:
$m_2 \sim M_{BZ}^3/m_{H}^2 \sim M_\L$;
see~\eqref{e:mH-BHM} and~\eqref{e:lBZ}.

Finally, the lightest (``electron'') neutrino mass is then
estimated by the (T-duality) seesaw formula
\be
m_1 \sim M_\L^2/m_2 \sim
M_\L \sqrt{\frac{M_\L}{m_{3}}}
\sim 10^{-4}\,\text{eV}. \label{e:m1}
\ee
According to the Particle Data Group\cite{rPDG22}, the sum of neutrino masses (coming from cosmology) is bounded by
$10^{-1}$\,eV, which is satisfied by the above normal hierarchy of neutrino masses. Incidentally, the above reasoning {\em predicts} the normal hierarchy, i.e., that $(m_3/m_2)<(m_2/m_1)$.
Also, these values satisfy the constraint on the square of the differences of masses, 
$(10^{-2}\text{--}10^{-5})\,\text{eV}^2$, coming from neutrino oscillation experiments. It would exciting to learn if these values for the neutrino masses are experimentally falsifiable.

All these estimates for quark lepton and Higgs masses and for the cosmological constant mass scale are upper bounds; this bound for $m_u$ and $m_d$ essentially being given by $M_{BZ}$. 
We thus expect an attractor mechanism (as in \cite{Argyriadis:2019fwb}) 
that would ``glue'' all these 
values to their upper bounds. This would be consistent with the existence of a moduli-free self-dual fixed point in metastring theory \cite{Freidel:2015pka} that could explain
the apparent criticality of the Standard Model parameters \cite{Froggatt:1995rt}.
Finally, all these bounds on the fermion masses, much as the bounds on the cosmological constant and the Higgs mass, are determined in terms of the Hubble and the Planck mass scales.

\subsection{Quantum Gravity and Fermion Mixing}
\label{s:fMixing}
We now comment on the CKM and PMNS mixing matrices, generally given in the format\cite{rPDG22}
\begin{equation}
\begin{pmatrix}
 c_{12}\,c_{13} & s_{12}\,c_{13} & ~s_{13}\,e^{-i\d}~\\
 -s_{12}\,c_{23}-c_{12}\,s_{13}\,s_{23}e^{i\d}
        & -c_{12}\,c_{23}-s_{12}\,s_{13}\,s_{23}e^{i\d}
                 & c_{13}\,s_{23}\\
  s_{12}\,s_{23}-c_{12}\,s_{13}\,c_{23}\,e^{i\d}
        & -c_{12}\,s_{23}-s_{12}\,s_{13}\,c_{23}\,e^{i\d}
                 & c_{13}\,c_{23}\\
\end{pmatrix},
 \label{e:mixMat}
\end{equation}
where $c_{ij}\define\cos(\theta_{ij})$ and $s_{ij}\define\sin(\theta_{ij})$, with $0\leqslant\theta_{ij}\leqslant\pi/2$ and $\delta=\delta_{13}$. In particular:
\begin{equation}
    V_{CKM}=
\begin{pmatrix}
 V_{ud} & V_{us} & V_{ub}\\
 V_{cd} & V_{cs} & V_{cb}\\
 V_{td} & V_{ts} & V_{tb}\\
\end{pmatrix}
=
\begin{pmatrix}
 0.97373 & 0.2243 & 0.00382\\
 0.221 & 0.975 & 0.0408\\
 0.0086 & 0.0415 & 1.014\\
\end{pmatrix},
 \label{e:mixCKM}
\end{equation}
with experimental errors in the last digits\cite{rPDG22}.

\paragraph{The CKM Matrix:}
The above cascading seesaw mass estimates also imply a similar structure for the fermion mixing matrices. To this end, we
write three key CKM matrix elements (which determine the three independent mixing angles) in terms of $M_{BZ}$ (as the relevant vacuum scale), and the quark masses as obtained above.
The estimates listed below consist of two seesaw factors
as motivated in the beginning of this section, and
agree very well with Bjorken's parametrization and values\cite{Bjorken:2013aa}
(given in parentheses):
\begin{alignat}9
|V_{cb}| &\sim \frac{M_{BZ}}{\sqrt{m_b\,m_d}}
&&\sim \sqrt{\frac{M_{BZ}}{{m_b}}} \sqrt{\frac{M_{BZ}}{{m_d}}}
 &&\sim 0.050 \quad (0.041),\qquad
 &&(\leadsto\theta_{23})
\label{e:Vcb}
\intertext{(essentially, $(M_{BZ}/m_b)^{1/4}$)
as well as}
|V_{td}| &\sim \frac{M_{BZ}}{\sqrt{m_b\,m_s}}
&&\sim \sqrt{\frac{M_{BZ}}{{m_b}}} \sqrt{\frac{M_{BZ}}{{m_s}}}
 &&\sim 0.011 \quad (0.008)
 &&(\leadsto\theta_{12})
\label{e:Vtd}
\intertext{(essentially, $(M_{BZ}/m_b)^{3/4}$)
and finally}
|V_{ub}| &\sim \frac{M_{BZ}}{\sqrt{m_b\,m_b}}
&&\sim \sqrt{\frac{M_{BZ}}{{m_b}}} \sqrt{\frac{M_{BZ}}{{m_b}}}
 &&\sim 0.002 \quad (0.003)
 &&(\leadsto\theta_{13})
\label{e:Vub}
\end{alignat}
(essentially, $M_{BZ}/m_b$).
All these estimates are of course expressible in terms of the $M_{CH}/M_P$ ratio of the two ultimate horizon-scales; see Table~\ref{t:Masses}.
To compare with~\eqref{e:mixMat} and~\eqref{e:mixCKM}: {\em first,} $\theta_{13}$ is determined from~\eqref{e:Vub}; with that, $\theta_{23}$ is determined from~\eqref{e:Vcb} {\em second,} and with those, $\theta_{12}$ is determined from~\eqref{e:Vtd}.
As in Bjorken's parametrization (using $(M_{BZ}/m_b)^{1/2}$ instead of~\eqref{e:Vcb}), 
these values are quite good when
compared to experiment, except perhaps for the
first value which here depends on the value of the down quark, and is, according to our prescription off by an approximate factor of 
$10$ from the observed value.

\paragraph{The PMNS Matrix:}
Neutrino mixing is parametrized much as the CKM matrix,~\eqref{e:Vcb}--\eqref{e:Vub}, but using $M_\L$ instead of $M_{BZ}$ as motivated in the above discussion of neutrino masses, as well as by replacing $m_b\to m_3$, $m_s\to m_2$ and $m_d\to m_1$, as should be evident.
We also take into account that $m_3$ is known up to a factor of $1/10$ in the above formula for the heaviest neutrino mass (the observed data\cite{Esteban:2020aa,rPDG22} is included in parentheses):
\be
|U_{\mu 3}| \sim \frac{M_\L}{\sqrt{m_3 m_1}} 
\sim \sqrt{\frac{M_{\Lambda}}{{m_3}}} \sqrt{\frac{M_{\Lambda}}{{m_1}}}
\sim 0.50, \quad (0.63)
\label{e:Um3}
\ee
(essentially, 
$(M_\L/m_3\overset{\sss\eqref{e:m3}}{\sim} M_{SM}/M_P)^{1/4}$) as well as
\be
|U_{\tau 1}| \sim \frac{M_\L}{\sqrt{m_3 m_2}} 
\sim \sqrt{\frac{M_{\Lambda}}{{m_3}}} \sqrt{\frac{M_{\Lambda}}{{m_2}}}
\sim 0.13, \quad (0.26) 
\label{e:Ut1}
\ee
(essentially, $(M_\L/m_3\sim  M_{SM}/M_P)^{3/4}$)
and finally
\be
|U_{e3}| \sim \frac{M_\L}{\sqrt{m_3 m_3}} 
\sim \sqrt{\frac{M_{\Lambda}}{{m_3}}} \sqrt{\frac{M_{\Lambda}}{{m_3}}}
\sim 0.06, \quad (0.14)
\label{e:Ue3}
\ee
(essentially, $M_\L/m_3\sim  M_{SM}/M_P$).
These values, are to first order, quite good when compared to
the observed data \cite{rPDG22}.
\footnote{Bjorken has different masses for neutrinos and his PMNS matrix is of the tri-bimaximal type \cite{rBJ-MM}. In his treatment of the neutrino sector the characteristic scale is still $M_{BZ}$.}

Although the numerical values of the CKM and PMNS matrices are quite different, we have shown that their underlying pattern is the same.
(However, we do not know the precise origin of this
underlying pattern. As emphasized above, that would require a fermionic analog of the Abel-Dienes stringy formula for the Higgs mass\cite{Abel:2021tyt}.)
The crucial difference stems 
from the appearance of $M_\L$ for neutrinos in place of $M_{BZ}$ for quarks and charged leptons.
Also, whereas the heaviest neutrino is determined by the Weinberg dimension 5 operator and the crucial $M_{BZ}\to M_\L$ replacement,
the other two neutrino masses follow the same pattern found in the case of quarks and charged leptons.
In our approach the CP violating phases would come from the SM calculation and the dual SM sector as
well from the intrinsic CP violation of quantum gravity in the non-perturbative metastring theory.

In conclusion to this section, note that the Standard Model of the observed kind 
(and not its SuSy extension) could be obtained by understanding the gauge groups as
general quantum phases. 
Recall that 
the $E_8$ prediction
of string theory as an overarching gauge group could be understood from the point of view of octo-octonionic geometry, which by dimensional reduction
to real-octonionic geometry
gives the (non-associative) geometry of the unique octonionic quantum theory captured by the octonionic projective geometry of quantum theory,
with the isometry group of $F_4/SO(9)$, whereas $SO(9)$ is the general quantum phase,
that upon its compatibility with the 4-dimensional Poincare group leads to the Standard Model gauge group\cite{Gunaydin:1974fb,Gunaydin:1973rs}.
This  is different from the usual GUT logic, but it points to a possible robustness of the Standard Model group
(and its dual Standard Model of the dark sector, coming from the other $E_8$ in heterotic string theory).
Note that this fits within the metastring formulation, because the heterotic string is constructed from the bosonic string in 26 dimensions\cite{Casher:1985ra},
and the metastring is just its T-duality covariant chiral (phase space-like) formulation, that, in general, allows for non-associative background geometry.

Finally, from this bottom-up point of view (our discussion has been in some
sense top-down) a modular quantization of the SM, coupled to the modular extension of general relativity, should give 
the structure that is implied by the top-down quantum gravity/string theory
(gravitized quantum theory) approach.

\section{Outlook: Quantum Gravity and Observation} 
\label{s:Coda}
In this paper we have presented a review of the new approach to quantum gravity based on gravitization of quantum theory, where by gravitized quantum theory we mean a formulation of quantum theory in which its geometry and topology of the space of states becomes fully dynamical.
This approach is based on the new calculation of the cosmological constant, $\L_{cc}$ based on the phase space reasoning and the holographic bound.
Gravitization of quantum theory can be experimentally probed by searching for higher order quantum interference phenomena in the presence of gravity, as
well as by probing the statistics of spacetime quanta via gravitational interferometry. By extending the computation of the cosmological constant, we have also evaluated the masses of the Higgs boson, as well as quarks and leptons and their mixing matrices.

Perhaps the most dramatic prediction of dynamical Born geometry implies ``gravitization of quantum theory''\cite{Freidel:2014qna,Berglund:2022qcc,Berglund:2022skk},and the presence of intrinsic and irreducible triple (and higher order) interference\cite{Sorkin:1994dt} 
in the presence of gravity\cite{Berglund:2023vrm}. This would be a new quantum probe of quantum spacetime and a new avenue
in quantum gravity phenomenology\cite{Hossenfelder:2012jw,Addazi:2021xuf}.
Similarly, the crucial seesaw formula, $\delta \sim \sqrt{l_{IR}\, l_{UV}}$, 
(with a characteristic IR length-scale $l_{IR}$ and the characteristic UV length-scale $l_{UV}$) found in the context of the computation of the vacuum energy, appears in other related contexts, such as the gravitational wave interferometry probes of quantum gravity; see, for example,\cite{Verlinde:2019xfb}. In that context, our vacuum energy calculation can be performed on the level of the causal diamond of the interferometer ($l_{IR}$ being given by the length of the interferometer and $l_{UV}$ by the Planck length), leading to the same seesaw formula, except interpreted as an empirical probe of modular spacetime. 
As we emphasized in Section~\ref{s:EspGQ}, the existing LIGO-VIRGO sensitivity is enough to test this seesaw relation for interferometers with $l$ of the order of a kilometer.

The geometric mean, seesaw formula for the associated length scale of the cosmological constant, $l_{cc}$~\eqref{e:lcc=llP}, exhibits UV/IR mixing, and that makes $l_{cc}$ radiatively stable and natural.
The logic of this resolution of the cosmological constant problem, with input from the Abel-Dienes stringy calculation, extends naturally to the Higgs mass~\eqref{e:mH-BHM}.
The same idea applies to the masses and mixing of quarks and leptons.
We have emphasized the quantum gravity nature of the cosmological constant,
the Higgs mass and the masses and mixing angles of quarks and leptons.
One important new ingredient in this reasoning is quantum contextuality (instead of the standard anthropic reasoning) which stems from the string/modular QFT vacuum being governed by Born geometry based on the modular phase space view of quantum spacetime \`a la\cite{Freidel:2022ryr}.
The crucial interplay of 
 ({\small\bf1})~phase space, 
 ({\small\bf2})~Born geometry, 
 ({\small\bf3})~the Bekenstein bound, 
 ({\small\bf4})~mixing between ultraviolet (UV) and infrared (IR) physics, and 
 ({\small\bf5})~modular invariance in string theory (in its intrinsically non-commutative, metastring formulation) was emphasized throughout this review.
All these are the essential features of the new view of quantum gravity as {\em gravitized quantum} (or {\em metaquantum}) theory.

Throughout the review we have repeatedly stressed the purely stringy or quantum-gravity-related effects which are fundamentally rooted in the properties of quantum spacetime.
Such effects are not part of the usual EFT lore, largely because EFT is defined in classical spacetime as a background.
This might sound disturbing given the success of EFT. Consequently, we have argued that EFT results, dressed up with holography, can be recovered in a singular limit of our computation of the vacuum energy. Given the fact that the usual compactification approach to string theory, and the associated string landscape and swampland\cite{Polchinski:2006gy,Agmon:2022thq,Berglund:2022qsb}, are closely tied to EFT, we conjecture that the application of holography in that context, and a seesaw relation between what are usually considered UV and IR cut-offs in EFT\cite{Cohen:1998zx}, could lead to a top-down realization of our computations and results, at a critical self-dual point (without moduli) which would hide the fundamental aspects of our discussion: modular spacetime, Born geometry and the metastring formulation.

The four-dimensional nature of our discussion may in this approach be related to the fundamental properties of strings at high temperature in the early universe\cite{Brandenberger:1988aj}. This could be then generalized to the computations of the Higgs mass and the masses and mixing matrices of quarks and leptons,
as discussed in this paper, revealing, perhaps, an attractor mechanism in string landscape and swampland. Similarly the observed cosmological background (de Sitter spacetime) which is usually problematic in string theory\cite{Berglund:2022qsb} (and thus leads to various conjectures in the
landscape\cite{Polchinski:2006gy} and swampland approaches\cite{Agmon:2022thq}) could be understood as the natural cosmological bound state of the 
non-perturbative matrix model formulation of metastring theory, and as such is a consequence of the view on quantum gravity as gravitized quantum theory.

In conclusion,
we list some further phenomenological implications of our work.

Our calculation of the cosmological constant introduces a new quantum number~\eqref{e:N}, $N$, which may be probed in gravitational waves, via gravitational wave ``echoes'': In particular, our result~\eqref{e:NllP} relates the number of phase space boxes to the Bekenstein bound, $N \sim l^2/l_P^2$. It can therefore be used for black holes, where $l\to l_{bh}$ is the size of the black hole horizon, where it is naturally related\cite{Bekenstein:1995ju}. In this case, the relevant quantization number, $N_{bh} \sim l_{bh}^2/l_P^2$, for black holes is of the order of $10^{80}$, and a possible observable feature of this quantization, $l_{bh}^2 \sim N l_P^2$, might be via the ``gravitational wave echoes'' \cite{Wang:2019rcf, Cardoso:2019apo} --- in the ``quantum chaos'' phase, given the enormous value of $N$. We also observe that in the context of the non-perturbative formulation of metastring theory via a gravitized matrix quantum theory discussed in section 3, black holes appear as natural astrophysical bound states of the fundamental partonic quantum spacetime degrees of freedom. Various observational astrophysical consequences of this picture are yet to be explored.

Furthermore, seesaw formulae for the SM fermion masses follow from the same reasoning that lead to the cosmological constant~\eqref{e:lcc=llP} and the Higgs mass~\eqref{e:mH-BHM} seesaw formulae. In that situation, a new Bjorken-Zeldovich scale can be deduced (by analogous reasoning) which enters into Bjorken-like seesaw formulae for all masses of charged elementary fermions.\footnote{In a fermionic \eqref{e:AbelDienes2}-like formula, the $\cX_\j$-insertion term must be proportional to gauge charges, and so is absent for neutrinos. In any explicit model-dependent calculation such as\cite{Faraggi:1991be}, the RG equations ``tie'' the heaviest charged fermion masses to the electroweak scale, while for neutrinos the relevant RG equations extend the UV scale to $M_{SM}\sim10^{15-16}$\,GeV.} This approach seems to proffer a new view on the observed three generations of quarks and leptons as well as their respective mixing matrices.
Here we point out an analogy with critical phenomena and the mean field/Landau-Ginzburg (LG) approach which gives ``square root type'' formulae, or the critical index of 1/2, without any anomalous dimensions, and which are, in turn, introduced by a more precise renormalization group (RG) treatment of the LG like description. In our case, the analogue of LG is the modular field theory extension of the SM and gravity. Our formulae should therefore be understood in the ``mean field theory sense''. In the context of modular field theory (a consistent limit of metastring theory) we also expect a {\em double\/} RG that is sensitive both to UV and IR scales\cite{Freidel:2017xsi}. Also, all these formulae can be rewritten ultimately using only the Hubble (IR) scale $M_{CH}$ and the Planck (UV) scale $M_P$. These are the only two scales that appear in all expressions for the cosmological constant, the Higgs mass and the masses and mixing of quarks and leptons.

We also point out that in the visible sector we ultimately have to work with
modular fields $\phi(x, \tilde x)$\cite{Freidel:2017nhg}. This is not so
in the Standard Model (SM) as it is understood at the moment,
but is implied by the modular polarization
and our argument about the bounds of fermion masses.
Thus, the modular SM fields should know about
the symplectic and also the biorthogonal 
structures associated with $x$ and $\tilde x$.
(This suggests a kind of generalized mirror symmetry in the visible sector.)
This is what induces two distinct seesaw formulae
(one non-commutative/symplectic, and one T-dual/biorthogonal),
naturally yielding three generations in $x$-spacetime 
(a heavy fermion and its two seesaw copies).
The invisible (dark) sector is spanned by the dual 
fields $\tilde \phi (x, \tilde x)$, which
may well be subject to a third
quantization indeterminacy because of an induced 
non-commutativity between visible $\phi$ and 
dual (invisible/dark) $\tilde\phi$ fields (fuzzy dark matter).
Thereby, while one may be able to deduce
the bounds on the parameters of the Standard Model (SM) in
string theory/quantum gravity, the ensuing indeterminacy in the parameters 
of the dual Standard Model (the dark sector)
should then be reciprocal to the relatively high precision (small indeterminacy) 
of the SM parameters.
(This would be in the spirit of the old ``third quantization'' proposal\cite{Strominger:1988si}). 

We stress that metaparticles\cite{Freidel:2018apz}
(zero modes of the metastring) represent a generic prediction of metastring theory and the dark matter sector can be seen as coming from a dual Standard Model with a dynamics that is entangled/correlated with the visible Standard Model\cite{rBHM10}.
The dark matter degrees of freedom are thus tied to the dual particles to the visible SM particles\cite{Berglund:2022qsb}.
Furthermore, this approach shows dark energy (modeled as the cosmological constant) to be the curvature of the dual spacetime, and naturally small\cite{Berglund:2022qsb}.
The natural relation between the dark matter and dark energy sectors in our formulation (see also \cite{Edmonds:2024qsj}), as well as the relation between the visible and dark sectors, offers, apart from quantum contextuality, a new view on the coincidence problem in cosmology\cite{Polchinski:2006gy}.

\paragraph{Acknowledgments:}
Many thanks to P.~Berglund, L.~Freidel, A.~Geraci, J.~Kowalski-Glikman, R.\,G.~Leigh and D.~Mattingly  
for insightful collaborations and illuminating discussions.
We also thank S.~Abel, N.~Afshordi, L.~Boyle, \v{C}.~Brukner,  T.~Curtright, 
K.~Dienes, M.~Dimitrijevi\'{c} \'{C}iri\'{c}, S.~Dimopoulos, 
P.~Draper, J.~Erlich, E.~Guendelman, L.~Hardy, Y.~H.~He, J.~Heckman, P.~Huber, N.~Ilic, T.~Jacobson, V.~Jejjala, M.~Johnson,
T.~Kephart, J.~Khoury, L.~Lehner, E.~Livine, A.~Mazumdar, L.~McAllister, H.~P\"{a}s, S.~Pasterski, M.~Pospelov, V.~Radovanovi\'{c}, P.~Ramond, M.~M.~Sheikh-Jabbari, M.~Shifman, 
D.~Stojkovic, T.~Takeuchi and N.~Turok for interesting questions and comments. 
Special thanks to Branko Dragovich of the Serbian Academy of Nonlinear Sciences for urging us to write this review.
TH is grateful to the Department of Physics, University of Maryland, and the Physics
Department of the University of Novi Sad, Serbia, for recurring hospitality and resources.
DM is supported by the  Julian  Schwinger Foundation and
the U.S. Department of Energy (under contract DE-SC0020262) and he thanks Perimeter Institute for hospitality and support.
DM is grateful to Dobrila and Aleksandra Minic and Esther and David Minic-Rosenthal for love and kindness.
DM wishes to thank Emil and Tanja Savic for generous support during a tremendously challenging time. DM dedicates this review to the timeless memory of his beloved wife Joy Rosenthal (1966--2023).

%%%%%%%%%%%%%%%%%%%%%%%%%%%%%%%%%%%%%%%%%%%%%%%%
%: References
\begingroup
%\clearpage
\raggedright
\small\baselineskip=13pt \parskip=0pt plus2pt minus1pt
\addcontentsline{toc}{section}{\numberline{}References}
%\bibliographystyle{utphys}
%\bibliography{dSinST}
%\endgroup
%\end{document}
%%%%%%%%%%%%%%%%%%%%%%%%%%%%%%%%%%%%%%%%%%%%%%%%

\endgroup
\end{document}